\documentclass[
twocolumn,
amsmath, amssymb, amsfonts,preprintnumbers,aps,prd,longbibliography
,bibnotes
]{revtex4-2}

\usepackage{graphicx}
\pdfoutput=1

\usepackage{dcolumn}% Align table columns on decimal point
\usepackage{bm}% bold math
 \usepackage{amstext}

 \usepackage{color}
 \usepackage{bbold}
 \usepackage{delimset} % for brackets
\usepackage[caption=false,justification=justified]{subfig}
\usepackage[colorlinks=true]{hyperref}
\usepackage{orcidlink}
\usepackage{float}
\usepackage[normalem]{ulem}
\usepackage{mathtools}

\usepackage{tikz}
\usetikzlibrary{decorations.pathmorphing}
\usetikzlibrary{decorations.markings}
\usetikzlibrary{positioning, shapes, snakes}
\usetikzlibrary{arrows.meta}

\usepackage{slashed}

\usepackage{amssymb}

\usepackage[]{shuffle}

\begin{document}

\author{Yong Zhang \!\orcidlink{0000-0002-3522-0885}}

\email{zhangyong1@nbu.edu.cn}

\author{Pengfei Zhu
\!\orcidlink{0009-0000-2310-947X}}

\email{2511690107@nbu.edu.cn}

\affiliation{
Institute of Fundamental Physics and Quantum Technology\\ \& School of Physical Science and Technology, Ningbo University, Ningbo, Zhejiang 315211, China }

\title{Fermionic hidden zeros}
\begin{abstract}
We uncover and prove a general class of hidden zeros in tree-level
amplitudes with massless fermions.  For arbitrary even numbers and arbitrary arrangements
of massless real adjoint fermions and gluons, color-ordered gauge-theory
amplitudes in $D=4,6,10$ vanish on general rectangular kinematic loci
supplemented by species-dependent bridge conditions.  The fermion--fermion
bridge is governed simply by the vector current
$\chi_i\gamma^\mu\chi_j$, while mixed and gluonic bridges are fixed by the
corresponding gauge-covariant contractions.  A worldsheet analysis proves
the result at arbitrary multiplicity and traces its universality to the
local fusion channels $ff\to g$, $fg/gf\to f$, and $gg\to g$, with no new
primitive bridge structures appearing at higher fermion multiplicity.
The same mechanism extends to ten-dimensional amplitudes with gravitons
and arbitrary even numbers of gravitinos in factorized gamma-traceless
polarizations, and to single-trace Einstein--Yang--Mills amplitudes with
adjoint gluinos and gravitons.
These results expose a finite local structure underlying fermionic hidden
zeros across gauge and gravitational amplitudes.
\end{abstract}

\maketitle

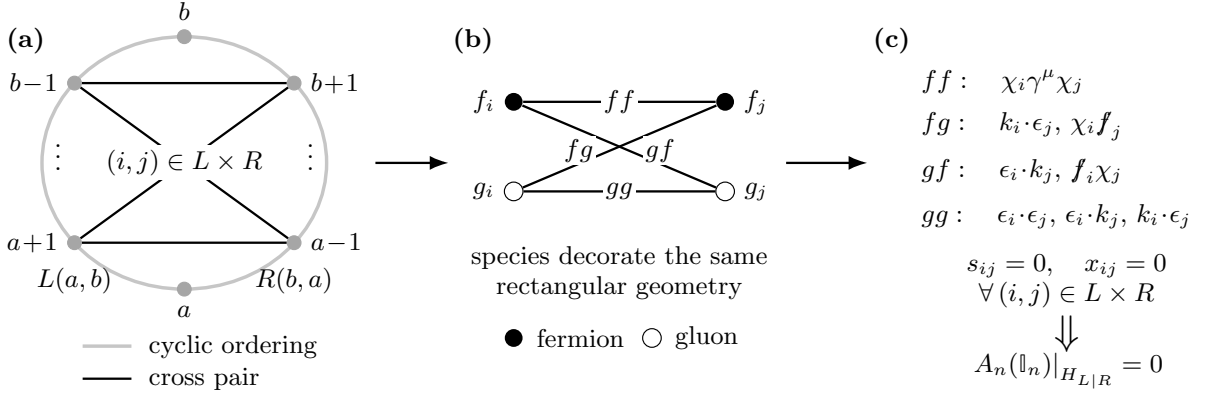
\begin{figure*}[t]
\centering
\resizebox{0.90\textwidth}{!}{
\begin{tikzpicture}[
    >=Latex,
    every node/.style={font=\small},
    state/.style={
        circle,
        draw=gray!70,
        fill=gray!70,
        inner sep=1.7pt
    },
    fdot/.style={
        circle,
        fill,
        inner sep=2.2pt
    },
    gdot/.style={
        circle,
        draw,
        fill=white,
        inner sep=2.3pt
    },
    arrow/.style={->,thick},
    bridge/.style={thin},
    active/.style={thick},
]

% ================================================================
% (a) Cyclic ordering and rectangular cross pairs
% ================================================================
\node[font=\bfseries] at (-0.45,2.8) {(a)};

% Cyclic ordering
\draw[
    gray!45,
    line width=1.2pt
]
(1.50,1.30) ellipse [x radius=1.75, y radius=1.55];

% ------------------------------------------------
% Two complementary arcs separated by a and b
% ------------------------------------------------

% Left arc: L(a,b)
\node[state,label=left:{$b\!-\!1$}] (lm) at (0.15,2.28) {};
\node at (-0.05,1.48) {$\vdots$};
\node[state,label=left:{$a\!+\!1$}] (l1) at (0.15,0.32) {};

% Distinguished leg b
\node[state,label=above:{$b$}] (spb) at (1.50,2.85) {};

% Right arc: R(b,a)
\node[state,label=right:{$b\!+\!1$}] (r1) at (2.85,2.28) {};
\node at (3.05,1.48) {$\vdots$};
\node[state,label=right:{$a\!-\!1$}] (rm) at (2.85,0.32) {};

% Distinguished leg a
\node[state,label=below:{$a$}] (spa) at (1.50,-0.25) {};

% Labels for the two arcs
\node at (0.15,-0.15) {$L(a,b)$};
\node at (2.85,-0.15) {$R(b,a)$};

% ------------------------------------------------
% Rectangular cross pairs
% ------------------------------------------------
\draw[active] (lm) -- (r1);
\draw[active] (lm) -- (rm);
\draw[active] (l1) -- (r1);
\draw[active] (l1) -- (rm);

\node[
    fill=white,
    inner sep=1.5pt
]
at (1.50,1.30)
{$(i,j)\in L\times R$};

% ------------------------------------------------
% Visual legend for panel (a)
% ------------------------------------------------
\draw[
    gray!45,
    line width=1.2pt
]
(0.25,-0.95) -- (0.85,-0.95);
\node[anchor=west] at (0.95,-0.95)
{cyclic ordering};

\draw[active]
(0.25,-1.35) -- (0.85,-1.35);
\node[anchor=west] at (0.95,-1.35)
{cross pair};

% Arrow (a) -> (b)
\draw[arrow] (3.85,1.30) -- (4.75,1.30);

% ================================================================
% (b) Species-decorated rectangle
% ================================================================
\node[font=\bfseries] at (5.05,2.8) {(b)};

% Left states
\node[fdot,label=left:{$f_i$}] (bf1) at (5.55,2.05) {};
\node[gdot,label=left:{$g_i$}] (bg1) at (5.55,0.95) {};

% Right states
\node[fdot,label=right:{$f_j$}] (bf2) at (8.15,2.05) {};
\node[gdot,label=right:{$g_j$}] (bg2) at (8.15,0.95) {};

% All four pair types
\draw[active] (bf1) -- (bf2);
\draw[active] (bf1) -- (bg2);
\draw[active] (bg1) -- (bf2);
\draw[active] (bg1) -- (bg2);

% Labels on edges
\node[fill=white,inner sep=1pt] at (6.85,2.05) {$ff$};
\node[fill=white,inner sep=1pt] at (6.35,1.45) {$fg$};
\node[fill=white,inner sep=1pt] at (7.35,1.45) {$gf$};
\node[fill=white,inner sep=1pt] at (6.85,0.95) {$gg$};

\node[align=center] at (6.85,-0.05)
{species decorate the same\\rectangular geometry};

% Legend for particle species
\node[fdot] at (5.55,-0.85) {};
\node[anchor=west] at (5.72,-0.85)
{fermion};

\node[gdot] at (7.25,-0.85) {};
\node[anchor=west] at (7.42,-0.85)
{gluon};

% Arrow (b) -> (c)
\draw[arrow] (8.90,1.30) -- (9.90,1.30);

% ================================================================
% (c) Species-dependent bridges and hidden zero
% ================================================================
\node[font=\bfseries] at (10.20,2.8) {(c)};

\node[anchor=west] at (10.45,2.30)
{$ff:\quad \chi_i\gamma^\mu\chi_j$};

\node[anchor=west] at (10.45,1.75)
{$fg:\quad
k_i\!\cdot\!\epsilon_j,\,
\chi_i\slashed f_j$};

\node[anchor=west] at (10.45,1.20)
{$gf:\quad
\epsilon_i\!\cdot\!k_j,\,
\slashed f_i\chi_j$};

\node[anchor=west] at (10.45,0.65)
{$gg:\quad
\epsilon_i\!\cdot\!\epsilon_j,\,
\epsilon_i\!\cdot\!k_j,\,
k_i\!\cdot\!\epsilon_j$};

\node[align=center] at (12.35,-0.15)
{$s_{ij}=0,\quad x_{ij}=0$\\
$\forall\,(i,j)\in L\times R$};

\node[font=\Large] at (12.35,-0.80)
{$\Downarrow$};

\node[font=\bfseries] at (12.35,-1.25)
{$\left.A_n(\mathbb I_n)\right|_{H_{L|R}}=0$};

\end{tikzpicture}
}
\caption{
Fermionic hidden zeros.
(a) Two distinguished legs $a$ and $b$ divide the physical cyclic
ordering into the complementary open arcs
$L=L(a,b)$ and $R=R(b,a)$.  The labels $a\pm1$ and $b\pm1$ are
understood cyclically modulo $n$.  The black chords represent the
cross pairs $(i,j)\in L\times R$.
(b) Assigning particle species decorates the same rectangular geometry:
filled and open circles denote fermions and gluons, respectively.
(c) Setting $s_{ij}=0$ together with the corresponding
species-dependent bridge data $x_{ij}=0$ for every cross pair defines
the locus $H_{L|R}$, on which
$\left.A_n(\mathbb I_n)\right|_{H_{L|R}}=0$ in $D=4,6,10$.
}
\label{fig:fermionic-hidden-zero}
\end{figure*}

\section*{Introduction}

Hidden zeros reveal an unexpected simplicity of scattering amplitudes on
special kinematic subspaces.  They belong to a broader class of recently
discovered splitting phenomena that occur away from ordinary physical
poles.  Smooth splitting was first found for scalar amplitudes, which
decompose into three semi-local currents on special Mandelstam subspaces
\cite{Cachazo:2021wsz}.  A different class of rectangular kinematic
configurations was subsequently shown to produce hidden zeros and
factorization near the zeros in scalar, pion, gluon, and string amplitudes
\cite{Arkani-Hamed:2023swr}; related zeros of dual-resonance amplitudes had
already appeared in early string theory \cite{DAdda:1971wcy}.  These
structures were later unified and extended by the 2-split construction,
in which amplitudes factorize into two currents under suitable kinematic
and polarization constraints
\cite{Cao:2024gln,Cao:2024qpp}.

The emerging hidden-zero structure has since been explored from several
complementary viewpoints, including double-copy and
Bern--Carrasco--Johansson (BCJ) relations
\cite{Bern:2008qj,Bern:2010ue,Bartsch:2024amu,Li:2024qfp}, bootstrap, locality, and uniqueness
arguments
\cite{Rodina:2024yfc,Zhou:2026isc,Rodina:2026yaq},
on-shell recursion \cite{Jones:2025rbv}, and direct Feynman-diagram
analyses \cite{Zhou:2024ddy}.  For Yang--Mills (YM) amplitudes, the
rectangular zeros are accompanied by polarization constraints
\cite{Arkani-Hamed:2023swr}, and can be embedded into more general
factorization formulas in which amplitudes reorganize into sums of gluings
of lower-point amplitudes
\cite{Guevara:2024nxd,Zhang:2024efe}.  These developments suggest that hidden
zeros are not accidental cancellations of particular low-point amplitudes,
but reflect a more systematic organization of kinematics and
external-state data.

This raises the question of whether there is a universal field-theory
structure underlying fermionic hidden zeros.  Fermions carry spinorial
rather than vectorial data, while fermion--fermion, fermion--gluon, and
gluon--gluon pairs probe different Lorentz structures.  Moreover,
amplitudes with many external fermions rapidly develop complicated spinor
contractions.  It is therefore far from obvious whether the same
rectangular geometry persists at arbitrary fermion multiplicity and for a
general rectangle, where several independent cross-pair degenerations may
occur simultaneously, or whether new structures and increasingly
complicated constraints must appear.

In this Letter we show that the fermionic extension is remarkably simple.
Tree-level color-ordered gauge-theory amplitudes with arbitrary even
numbers and arbitrary arrangements of massless real adjoint fermions and
gluons possess a universal family of hidden zeros in $D=4,6,10$.  The
rectangular kinematic geometry is unchanged from pure YM; only the local
bridge associated with each cross pair depends on the particle species.
Most notably, the complete fermion--fermion bridge is governed by the
vector current
\begin{equation}
\chi_i\gamma^\mu\chi_j=0 .
\label{eq:intro-ff-bridge}
\end{equation}
Fermion--gluon and gluon--gluon pairs are controlled by similarly simple
gauge-covariant contractions.  Imposing the corresponding bridge data
across the rectangle makes the amplitude vanish for arbitrary
multiplicity, arbitrary fermion--gluon arrangements, and general
rectangular kinematics.

The simplicity of this result has a local worldsheet origin.  Using the
Ramond--Neveu--Schwarz (RNS) ambitwistor representation and the
Cachazo--He--Yuan (CHY) scattering-equation description of tree amplitudes
\cite{Mason:2013sva,Cachazo:2013gna,Cachazo:2013hca,Cachazo:2013iea},
we show that the relevant singular data close on the elementary fusion
channels
$ff\to g$, $fg/gf\to f$, and $gg\to g$.
Their coefficients are precisely the species-dependent bridge data, and no
new primitive local structures appear at higher fermion multiplicity.
Combined with the known degeneration structure of scattering equations on
rectangular kinematic loci \cite{Zhang:2024efe}, this gives an
all-multiplicity proof for general rectangles.

Beyond the Majorana-type adjoint theory used to formulate the main theorem,
the same local bridge structure can be inherited or reorganized in more
general fermionic matter sectors, including adjoint Dirac fermions and
fundamental quarks.  Simple one-line massive extensions are also discussed
in the Supplemental Material, alongside recent investigations of hidden
zeros in massive field theories \cite{CarrilloGonzalez:2026lnu}.
The same mechanism also extends to ten-dimensional
graviton--gravitino amplitudes in factorized gamma-traceless
polarizations, and to single-trace Einstein--Yang--Mills (EYM) amplitudes with
adjoint gluinos and gravitons.
These extensions indicate that the finite local fusion structure
underlying the fermionic hidden zeros persists more broadly across gauge
and gravitational amplitudes.

\section*{Fermionic hidden zeros}
\label{sec:fermionic-hidden-zeros}

We consider YM theory minimally coupled to adjoint fermionic matter,
\begin{align}
\mathcal L
&=
-\frac14 F_{\mu\nu}^a F^{a\,\mu\nu}
+\frac{i}{2}\bar\chi^a\gamma^\mu D_\mu\chi^a ,
\nonumber\\
D_\mu\chi^a
&=
\partial_\mu\chi^a+g f^{abc}A_\mu^b\chi^c .
\label{eq:lagrangian}
\end{align}
In $D=4,6,10$ we take a single massless real adjoint fermion species
in the corresponding spinor representation,\footnote{We use
Majorana spinors in $D=4$, symplectic Majorana--Weyl spinors in $D=6$,
and Majorana--Weyl spinors in $D=10$.  Fermions and antifermions are
therefore not introduced as independent external species, and all
fermionic legs are denoted by $f$.  Dimension-specific conventions are
summarized in the Supplemental Material.}
and consider tree-level color-ordered amplitudes with arbitrary even
fermion number and arbitrary numbers of gluons.  All external momenta
are taken outgoing.

For a physical cyclic ordering $\mathbb I_n$, choose two nonadjacent
distinguished legs $a$ and $b$.  With respect to the chosen cyclic
orientation, let $L(a,b)$ and $R(b,a)$ denote the two complementary
nonempty ordered sets of legs lying strictly between them.  Thus
\begin{equation}
\mathbb I_n=(a,L,b,R),
\qquad
L=L(a,b),
\qquad
R=R(b,a),
\label{eq:cyclic-LR-decomposition}
\end{equation}
and the associated rectangular kinematic locus is
\begin{equation}
s_{ij}=0,
\qquad
(i,j)\in L\times R,
\qquad
s_{ij}:=k_i\cdot k_j .
\label{eq:rectangular-kinematics}
\end{equation}
This construction is illustrated in
Fig.~\ref{fig:fermionic-hidden-zero}(a).

For explicit formulas we use the canonical representative
$\mathbb I_n=(1,2,\ldots,n)$ with
$
a=n,
b=m+1,
$
so that
\begin{equation}
L_m=\{1,\ldots,m\},
\,
R_m=\{m+2,\ldots,n-1\},
\,
1\leq m\leq n-3 .
\label{eq:LR}
\end{equation}

The fermionic extension is encoded entirely in the bridge associated
with each cross pair.  For a gluon we define
\begin{equation}
f_i^{\mu\nu}
=
k_i^\mu\epsilon_i^\nu-k_i^\nu\epsilon_i^\mu,
\qquad
\slashed f_i
=
\frac14 f_{i,\mu\nu}\gamma^{\mu\nu},
\end{equation}
while $\slashed k_i\chi_i=0$ for an external fermion.  The
species-dependent bridge data are
\begin{equation}
\boxed{
x_{ij}=
\begin{cases}
\left\{\chi_i\gamma^\mu\chi_j\right\},
& i_fj_f,
\\[1mm]
\left\{k_i\cdot\epsilon_j,\,
       \chi_i\slashed f_j\right\},
& i_fj_g,
\\[1mm]
\left\{\epsilon_i\cdot k_j,\,
       \slashed f_i\chi_j\right\},
& i_gj_f,
\\[1mm]
\left\{\epsilon_i\cdot\epsilon_j,\,
       \epsilon_i\cdot k_j,\,
       k_i\cdot\epsilon_j\right\},
& i_gj_g .
\end{cases}}
\label{eq:species-bridge}
\end{equation}
Here $x_{ij}=0$ means that every entry in the corresponding set
vanishes.  The same rectangular geometry is therefore decorated locally
according to the species of each cross pair, as summarized in
Fig.~\ref{fig:fermionic-hidden-zero}(b,c).

Defining
\begin{equation}
H_{L|R}
:=
\left\{
s_{ij}=0,\ x_{ij}=0
\ \middle|\
(i,j)\in L\times R
\right\},
\label{eq:HLR}
\end{equation}
our main result is
\begin{equation}
\boxed{
\left.
A_n(\mathbb I_n)
\right|_{H_{L|R}}
=0,
\qquad
D=4,6,10 .
}
\label{eq:main-hidden-zero}
\end{equation}
This holds for arbitrary multiplicity, arbitrary even fermion number,
arbitrary fermion--gluon ordering, and any allowed choice of
distinguished legs $a,b$.  The statement is understood at generic points
of the rectangular locus away from ordinary planar factorization poles.

We have checked Eq.~\eqref{eq:main-hidden-zero} directly at low
multiplicity, including an explicit ten-dimensional eight-point
eight-fermion calculation.  Representative low-point component checks are collected in the
Supplemental Material.  In the next section we explain the worldsheet
origin of the bridge conditions and prove
Eq.~\eqref{eq:main-hidden-zero} at arbitrary multiplicity and for a
general rectangle.

The Majorana-type formulation is not essential to the one-line sector.
Oriented Dirac and fundamental-matter versions, together with the
corresponding charge-sector relations for several Dirac lines, are
summarized in the Supplemental Material.

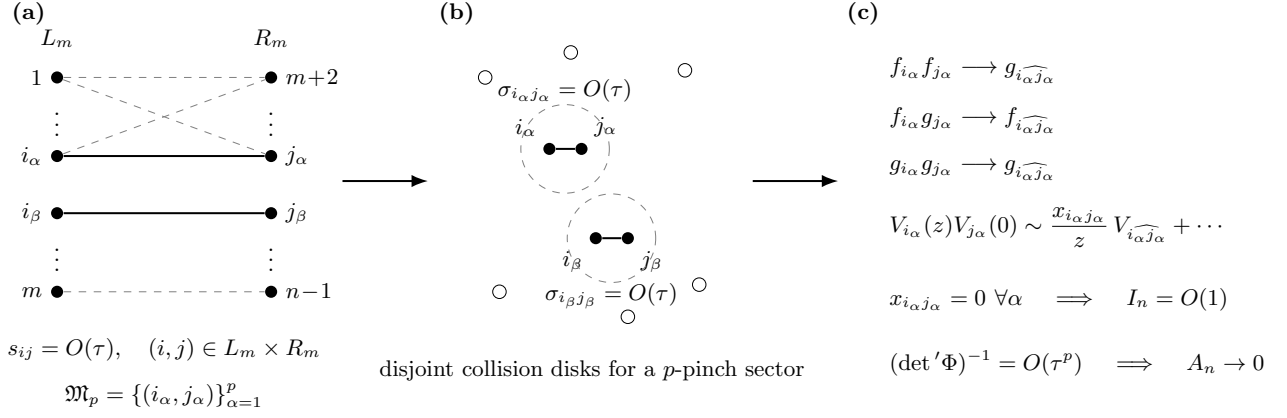
\begin{figure*}[t]
\centering
\resizebox{0.95\textwidth}{!}{
\begin{tikzpicture}[
    >=Latex,
    every node/.style={font=\small},
    dot/.style={circle,fill,inner sep=1.7pt},
    open/.style={circle,draw,fill=white,inner sep=1.9pt},
    arrow/.style={->,thick},
    guide/.style={dashed,gray},
    active/.style={thick},
]

% ================================================================
% (a) General rectangular locus and p-pinch matching
% ================================================================
\node[font=\bfseries] at (-0.4,2.9) {(a)};

% Left set: L_m
\node[dot,label=left:{$1$}] (L1) at (0,2.0) {};
\node at (0,1.45) {$\vdots$};
\node[dot,label=left:{$i_\alpha$}] (LiA) at (0,0.90) {};
\node[dot,label=left:{$i_\beta$}] (LiB) at (0,0.10) {};
\node at (0,-0.45) {$\vdots$};
\node[dot,label=left:{$m$}] (Lm) at (0,-1.0) {};

% Right set: R_m
\node[dot,label=right:{$m\!+\!2$}] (R1) at (3.0,2.0) {};
\node at (3.0,1.45) {$\vdots$};
\node[dot,label=right:{$j_\alpha$}] (RjA) at (3.0,0.90) {};
\node[dot,label=right:{$j_\beta$}] (RjB) at (3.0,0.10) {};
\node at (3.0,-0.45) {$\vdots$};
\node[dot,label=right:{$n\!-\!1$}] (Rn1) at (3.0,-1.0) {};

% Representative cross pairs
\draw[guide] (L1) -- (R1);
\draw[guide] (L1) -- (RjA);
\draw[guide] (LiA) -- (R1);
\draw[guide] (Lm) -- (Rn1);

% Active p-pinch matching
\draw[active] (LiA) -- (RjA);
\draw[active] (LiB) -- (RjB);

% Labels for the two sets
\node at (0,2.55) {$L_m$};
\node at (3.0,2.55) {$R_m$};

% Kinematic locus and matching
\node at (1.5,-1.80)
{$s_{ij}=O(\tau),\quad (i,j)\in L_m\times R_m$};

\node at (1.5,-2.45)
{$\mathfrak M_p=\{(i_\alpha,j_\alpha)\}_{\alpha=1}^{p}$};

% Arrow (a) -> (b)
\draw[arrow] (4.0,0.55) -- (5.2,0.55);

% ================================================================
% (b) Multiple worldsheet pinches
% ================================================================
\node[font=\bfseries] at (5.6,2.9) {(b)};

% Spectator punctures
\node[open] at (6.0,2.0) {};
\node[open] at (8.8,2.1) {};
\node[open] at (9.0,-0.9) {};
\node[open] at (6.2,-1.0) {};
\node[open] at (7.2,2.35) {};
\node[open] at (8.0,-1.35) {};

% First active pair
\node[dot,label=above left:{$i_\alpha$}] (Pa1) at (6.9,1.0) {};
\node[dot,label=above right:{$j_\alpha$}] (Pa2) at (7.35,1.0) {};
\draw[active] (Pa1) -- (Pa2);
\draw[guide] (7.12,1.0) circle (0.62);

% Second active pair
\node[dot,label=below left:{$i_\beta$}] (Pb1) at (7.55,-0.25) {};
\node[dot,label=below right:{$j_\beta$}] (Pb2) at (8.00,-0.25) {};
\draw[active] (Pb1) -- (Pb2);
\draw[guide] (7.77,-0.25) circle (0.62);

\node at (7.12,1.82)
{$\sigma_{i_\alpha j_\alpha}=O(\tau)$};

\node at (7.77,-1.05)
{$\sigma_{i_\beta j_\beta}=O(\tau)$};

\node at (7.5,-2.10)
{disjoint collision disks for a $p$-pinch sector};

% Arrow (b) -> (c)
\draw[arrow] (9.75,0.55) -- (10.95,0.55);

% ================================================================
% (c) Local fusion and global consequence
% ================================================================
\node[font=\bfseries] at (11.3,2.9) {(c)};

\node[anchor=west] at (11.55,2.10)
{$f_{i_\alpha}f_{j_\alpha}
\longrightarrow
g_{\widehat{i_\alpha j_\alpha}}$};

\node[anchor=west] at (11.55,1.40)
{$f_{i_\alpha}g_{j_\alpha}
\longrightarrow
f_{\widehat{i_\alpha j_\alpha}}$};

\node[anchor=west] at (11.55,0.70)
{$g_{i_\alpha}g_{j_\alpha}
\longrightarrow
g_{\widehat{i_\alpha j_\alpha}}$};

\node[anchor=west] at (11.55,-0.10)
{$V_{i_\alpha}(z)V_{j_\alpha}(0)
\sim
\dfrac{x_{i_\alpha j_\alpha}}{z}\,
V_{\widehat{i_\alpha j_\alpha}}
+\cdots$};

\node[anchor=west,font=\bfseries] at (11.55,-1.10)
{$x_{i_\alpha j_\alpha}=0\ \forall\alpha
\quad\Longrightarrow\quad
I_n=O(1)$};

\node[anchor=west,font=\bfseries] at (11.55,-2.00)
{$(\det{}'\Phi)^{-1}=O(\tau^p)
\quad\Longrightarrow\quad
A_n\to0$};

\end{tikzpicture}
}
\caption{
Worldsheet mechanism for the general fermionic hidden zero.
(a) On a general rectangular kinematic locus,
$s_{ij}=O(\tau)$ for all $(i,j)\in L_m\times R_m$.
A singular branch selects a $p$-pinch matching
$\mathfrak M_p=\{(i_\alpha,j_\alpha)\}_{\alpha=1}^{p}$ of mutually
row-- and column--disjoint cross pairs.
(b) On the corresponding singular branch, each matched pair produces a
worldsheet pinch $\sigma_{i_\alpha j_\alpha}=O(\tau)$, and the collision
regions are mutually disjoint at leading order.
(c) Each local collision closes on $ff\to g$, $fg/gf\to f$, and
$gg\to g$, with the singular fusion terms controlled by the corresponding
bridge data.  Schematically,
$V_i(z)V_j(0)\sim (x_{ij}/z)V_{\widehat{ij}}+\cdots$.
For the gluonic channel the schematic fusion is understood with the
regulated scaling $s_{ij}\sim z\sim\tau$, as explained in the text.
Imposing the bridges for all matched pairs therefore gives $I_n=O(1)$.
Together with $(\det{}'\Phi)^{-1}=O(\tau^p)$, this yields the hidden zero.
}
\label{fig:general-fusion}
\end{figure*}

\section*{Worldsheet origin and proof}
\label{sec:worldsheet-proof}

We now explain the worldsheet origin of the species-dependent bridge
conditions and prove the general rectangular hidden-zero theorem.  The
natural framework is the RNS ambitwistor string and its
scattering-equation representation of tree-level gauge-theory amplitudes
\cite{Mason:2013sva,Cachazo:2013gna,Cachazo:2013hca,
Cachazo:2013iea,Edison:2020uzf}.  Although the local Ramond and NS
operator algebra closely parallels conventional RNS language, the
worldsheet theory used here is the ambitwistor string, with scattering
equations replacing the finite-$\alpha'$ string moduli integrals.

The proof combines two ingredients.  The first is the known degeneration
structure of the scattering equations on rectangular kinematic loci,
which organizes singular solutions into mutually row-- and
column--disjoint cross-pair pinches \cite{Zhang:2024efe}.  The second is
the local RNS fusion algebra of the colliding external states.  We show
that its potentially divergent local data close on the three channels
\begin{equation}
\label{eq:local-fusion-summary}
ff\to g,
\qquad
fg/gf\to f,
\qquad
gg\to g,
\end{equation}
and are controlled by the bridge data introduced in
Eq.~\eqref{eq:species-bridge}.  For a local collision coordinate $z$, the
singular fusion terms take the schematic form
\begin{equation}
V_i(z)V_j(0)
\sim
\frac{x_{ij}}{z}\,
V_{\widehat{ij}}(0)
+\cdots ,
\label{eq:schematic-local-bridge-fusion}
\end{equation}
where the notation is schematic for the multi-component mixed and gluonic
bridges, and the gluonic channel is understood in the regulated sense
explained below.  Together these two ingredients are sufficient to
establish Eq.~\eqref{eq:main-hidden-zero} for arbitrary multiplicity and
a general rectangle.  The general mechanism is summarized in
Fig.~\ref{fig:general-fusion}.

\paragraph*{General rectangular degeneration.}

We work with the canonical rectangle \eqref{eq:LR},
and approach its Mandelstam locus as
\begin{equation}
s_{ij}
=
\tau\,\widehat s_{ij},
\qquad
(i,j)\in L_m\times R_m,
\qquad
\tau\to0,
\label{eq:general-rectangle-scaling}
\end{equation}
while keeping all ordinary planar factorization channels nonzero.

On this rectangular locus, the scattering-equation solutions are known
to organize into singular sectors characterized by mutually row-- and
column--disjoint cross-pair pinches
\cite{Zhang:2024efe}.  A $p$-pinch sector is specified by a matching
\begin{equation}
\mathfrak M_p
=
\left\{
(i_\alpha,j_\alpha)
\right\}_{\alpha=1}^{p}
\subset L_m\times R_m,
\label{eq:p-pinch-matching}
\end{equation}
with
\begin{equation}
i_\alpha\neq i_\beta,
\qquad
j_\alpha\neq j_\beta,
\qquad
\alpha\neq\beta,
\end{equation}
and
\begin{equation}
1\leq p\leq
\min\!\left(|L_m|,|R_m|\right).
\label{eq:p-pinch-range}
\end{equation}
The corresponding worldsheet separations scale as
\begin{equation}
\sigma_{i_\alpha j_\alpha}=O(\tau),
\qquad
\alpha=1,\ldots,p,
\label{eq:p-pinch-scaling}
\end{equation}
while the remaining separations stay finite at leading order.  For such
a $p$-pinch branch, the universal scattering-equation Jacobian obeys
\cite{Zhang:2024efe}
\begin{equation}
(\det{}'\Phi)^{-1}=O(\tau^p).
\label{eq:p-pinch-jacobian}
\end{equation}

\paragraph*{Local fusion algebra.}

The new fermionic input is local.  Consider one pinched pair
$(i_\alpha,j_\alpha)\in\mathfrak M_p$ and denote its collision coordinate by
\begin{equation}
z_\alpha
:=
\sigma_{i_\alpha j_\alpha}
\to0 .
\end{equation}
All remaining insertions may be collected into a spectator operator
$\mathcal X_\alpha$, so locally
\begin{equation}
I_n
=
\left\langle
V_{i_\alpha}(z_\alpha)
V_{j_\alpha}(0)\,
\mathcal X_\alpha
\right\rangle .
\end{equation}
The singular behavior in $z_\alpha$ is therefore determined entirely by
the OPE of the two colliding external states.

For two fermions, using the Ramond vertex
\begin{equation}
V_f^{(-1/2)}(\chi,k)
=
2^{-1/4}\chi^\alpha S_\alpha
e^{-\phi/2}e^{ik\cdot X},
\end{equation}
the leading spin-field OPE gives
\begin{equation}
V_{f,i}^{(-1/2)}(z)
V_{f,j}^{(-1/2)}(0)
=
\frac{1}{2z}
\left(\chi_i\gamma^\mu\chi_j\right)
V_{\widehat{ij},g,\mu}^{(-1)}(0)
+O(1).
\label{eq:ff-local-fusion-main}
\end{equation}
On $s_{ij}=0$, the fused momentum $k_i+k_j$ is null, so the singular term
is the massless fusion channel $ff\to g$.  Related gluino--gluino fusion
into a gluon in the superstring scaffolding framework was discussed in
Ref.~\cite{Chang:2025cqe}.  The fermion--fermion bridge
$\chi_i\gamma^\mu\chi_j=0$ therefore removes its complete principal part.
Equation~\eqref{eq:ff-local-fusion-main} is the explicit fermionic
realization of the schematic bridge-controlled fusion
\eqref{eq:schematic-local-bridge-fusion}.

The mixed channels have the same local structure: their potentially
divergent fusion coefficients are precisely the corresponding
multi-component bridge data in Eq.~\eqref{eq:species-bridge}.  The
gluon--gluon channel requires a slightly more refined statement.
Its bridge-controlled singular terms may carry additional factors of
$s_{ij}/\sigma_{ij}$, and terms proportional to
$s_{ij}/\sigma_{ij}$ may remain after the bridge conditions are imposed.
On the regulated degeneration
\begin{equation}
s_{ij}=O(\tau),
\qquad
\sigma_{ij}=O(\tau),
\end{equation}
such terms are only $O(1)$.  Thus Eq.~\eqref{eq:schematic-local-bridge-fusion}
captures the common fusion structure, while the gluonic channel is
understood with this regulated power counting.

Consequently, for every cross pair one has the uniform regulated statement
\begin{equation}
\left.
I_n
\right|_{
s_{ij}=O(\tau),\,
x_{ij}=0,\,
\sigma_{ij}=O(\tau)
}
=
O(1).
\label{eq:bridge-local-regularity}
\end{equation}
Equivalently, the bridge conditions remove all terms with negative total
$\tau$-scaling in the local collision expansion.  This statement is
independent of the number and species of the spectator states; in
particular, picture changing at arbitrary Ramond multiplicity introduces
no additional primitive divergent data.  Details are given in the
Supplemental Material.

\paragraph*{General multi-pinch proof.}

We now return to an arbitrary $p$-pinch sector
$\mathfrak M_p$.  Since the pinched pairs are mutually row-- and
column--disjoint, their collision regions are disjoint at leading order.
The regulated local statement
\eqref{eq:bridge-local-regularity} therefore applies independently to
every pair in $\mathfrak M_p$.  Imposing the full bridge conditions on
$H_{L_m|R_m}$ removes all negative total $\tau$-scaling from the joint
collision expansion, and hence
\begin{equation}
\left.
I_n
\right|_{\mathfrak M_p,H_{L_m|R_m}}
=
O(1).
\label{eq:multipinch-half-integrand-main}
\end{equation}

Moreover, no Parke--Taylor edge directly connects a leg in $L_m$ to a
leg in $R_m$ for the canonical ordering, so
\begin{equation}
\operatorname{PT}(\mathbb I_n)=O(1)
\label{eq:multipinch-pt-main}
\end{equation}
throughout the degeneration.  Combining
Eqs.~\eqref{eq:p-pinch-jacobian},
\eqref{eq:multipinch-half-integrand-main}, and
\eqref{eq:multipinch-pt-main}, each solution in the $p$-pinch sector
contributes
\begin{equation}
\left.
\frac{
\operatorname{PT}(\mathbb I_n)\,I_n
}{
\det{}'\Phi
}
\right|_{\mathfrak M_p,H_{L_m|R_m}}
=
O(\tau^p)
\longrightarrow0 .
\label{eq:multipinch-suppression-main}
\end{equation}

At a generic point of the rectangular locus away from ordinary planar
factorization poles, there are no regular solution branches
\cite{Zhang:2024efe}, while every singular branch is described by one
of the matchings $\mathfrak M_p$.  Summing over all allowed matchings
and all
$1\leq p\leq\min(|L_m|,|R_m|)$
therefore reproduces Eq.~\eqref{eq:main-hidden-zero}.
This completes the proof of the general fermionic hidden-zero theorem.

\section*{Hidden zeros with gravitons}
\label{sec:gravity-extensions}

The same local mechanism extends directly to gravitational states at the
level of the factorized CHY integrand.  Once the appropriate bridge
conditions are imposed, the relevant RNS half-integrands remain
$O(1)$ on every multi-pinch sector of a general rectangular degeneration.
The general scattering-equation argument of the previous section then
gives a hidden zero whenever the remaining CHY factors are regular.  We
illustrate this for ten-dimensional graviton--gravitino amplitudes and
for single-trace EYM amplitudes with adjoint gluinos and gravitons.  No new local fermionic fusion data are required.

\medskip
\paragraph*{Gravitons and gravitinos.}

Consider first ten-dimensional gravity.  Choose two distinguished legs
$a,b$ and partition the remaining external states into two nonempty
sets,
\begin{equation}
L\sqcup R
=
\{1,\ldots,n\}\setminus\{a,b\},
\qquad
L,R\neq\varnothing .
\label{eq:gravity-partition}
\end{equation}
Unlike the color-ordered gauge-theory case, no ordering is attached to
$L$ or $R$.  The rectangular kinematics is simply
$s_{ij}=0$ for all $(i,j)\in L\times R$.

The gravity amplitude is represented by two RNS half-integrands,
\begin{equation}
\mathcal M_n
=
\int d\mu_n\,
I_n\,\widetilde I_n .
\label{eq:gravity-chy}
\end{equation}
For a fixed gravitino sector we use the factorized states
\begin{equation}
h:\quad
\varepsilon_{\mu\nu}
=
\epsilon_\mu\widetilde\epsilon_\nu,
\qquad
\psi_{3/2}:\quad
\Psi_{\mu\alpha}
=
\epsilon_\mu\widetilde\chi_\alpha,
\qquad
\slashed{\epsilon}\,\widetilde\chi=0 .
\label{eq:gravity-factorized-states}
\end{equation}
The last condition imposes the Rarita--Schwinger gamma-trace constraint
$\gamma^\mu\Psi_{\mu}=0$, projecting onto the spin-\(3/2\) sector.   Thus an amplitude with $2r$ such gravitinos is
purely bosonic in $I_n$, while $\widetilde I_n$ contains $2r$ fermionic
states.

For each cross pair, the two copies inherit the gauge-theory bridge
types
\begin{equation}
\begin{array}{c|cc}
\text{pair} & I_n & \widetilde I_n \\ \hline
hh          & gg & gg \\
\psi h      & gg & fg \\
h\psi       & gg & gf \\
\psi\psi    & gg & ff
\end{array}
\label{eq:gravity-bridge-table}
\end{equation}
with corresponding bridge data $x_{ij}$ and $\widetilde x_{ij}$.
Here $\widetilde x_{ij}$ denotes the same gauge-theory bridge
assignment as $x_{ij}$, but evaluated on the tilded polarization and
spinor data of the second half-integrand.  Imposing
$x_{ij}=\widetilde x_{ij}=0$ removes all contributions with negative
total $\tau$-scaling from both half-integrands on every pinched cross
pair.  At generic kinematics away from ordinary factorization poles, the
general multi-pinch argument of the previous section therefore gives
\begin{equation}
\begin{gathered}
s_{ij}=0,\qquad
x_{ij}=\widetilde x_{ij}=0,
\qquad
\forall\,(i,j)\in L\times R,
\\[1mm]
\Longrightarrow\qquad
\mathcal M_n
\bigl(h^{\,n-2r},\psi_{3/2}^{\,2r}\bigr)=0 .
\end{gathered}
\label{eq:gravitino-hidden-zero}
\end{equation}

\medskip
\paragraph*{Gravitons and gluinos.}

We next consider the single-trace EYM component
\begin{equation}
A_n^{\rm EYM}
\bigl(f^{\,2r};h^{\,n-2r}\bigr)
\label{eq:eym-component}
\end{equation}
with adjoint gluinos and gravitons.  We use
\begin{equation}
f:\ \chi,
\qquad
h:\ \epsilon_\mu\widetilde\epsilon_\nu ,
\label{eq:eym-states}
\end{equation}
for which the CHY integrand has the schematic form
\cite{Cachazo:2014xea}
\begin{equation}
A_n^{\rm EYM}
=
\int d\mu_n\,
\operatorname{PT}(\mathbb I_{\rm c})\,
I_n\,
\widetilde I_{\rm grav}(\mathcal H).
\label{eq:eym-chy}
\end{equation}
Here $\mathbb I_{\rm c}$ denotes the single color trace and
$\mathcal H$ the set of gravitons.  The half-integrand $I_n$ carries
the untilded $(\chi,\epsilon)$ data, while
$\widetilde I_{\rm grav}$ contains only the tilded graviton
polarizations.

In $I_n$, a gluino behaves locally as a fermion and a graviton through
its untilded polarization as a gluon.  Thus the physical pairs
$ff$, $fh$, $hf$, and $hh$ inherit respectively the
$ff$, $fg$, $gf$, and $gg$ bridges of
Eq.~\eqref{eq:species-bridge}.

The second half-integrand contains only the tilded graviton data, with
bridge conditions
\begin{equation}
\widetilde x_{ij}
=
\begin{cases}
\varnothing,
& i_fj_f,
\\[1mm]
\left\{k_i\cdot\widetilde\epsilon_j\right\},
& i_fj_h,
\\[1mm]
\left\{\widetilde\epsilon_i\cdot k_j\right\},
& i_hj_f,
\\[1mm]
\left\{
\widetilde\epsilon_i\cdot\widetilde\epsilon_j,\,
\widetilde\epsilon_i\cdot k_j,\,
k_i\cdot\widetilde\epsilon_j
\right\},
& i_hj_h .
\end{cases}
\label{eq:eym-tilde-bridge}
\end{equation}
Thus $\widetilde x_{ij}$ is simply the restriction of the usual
bosonic bridge to the polarization data present in
$\widetilde I_{\rm grav}$; for an $ff$ pair no additional condition
is required.

The only new ingredient is the color trace.  The partition
$(a,b;L,R)$ must be chosen so that no Parke--Taylor edge directly
connects $L$ and $R$, or equivalently every $L$--$R$ transition along
$\mathbb I_{\rm c}$ passes through $a$ or $b$.  Then
\begin{equation}
\operatorname{PT}(\mathbb I_{\rm c})=O(1)
\label{eq:eym-pt-regular}
\end{equation}
throughout the rectangular degeneration.  Imposing
$s_{ij}=0$ and $x_{ij}=\widetilde x_{ij}=0$ for every
$(i,j)\in L\times R$, the same general multi-pinch argument gives
\begin{equation}
\begin{gathered}
s_{ij}=0,\qquad
x_{ij}=\widetilde x_{ij}=0,
\qquad
\forall\,(i,j)\in L\times R,
\\[1mm]
\Longrightarrow\qquad
A_n^{\rm EYM}
\bigl(f^{\,2r};h^{\,n-2r}\bigr)=0 .
\end{gathered}
\label{eq:eym-hidden-zero}
\end{equation}

Thus gravity and single-trace EYM inherit the same local fermionic
bridge algebra.  Gravity requires only the two-copy bridge conditions,
while the color trace in EYM adds the requirement of Parke--Taylor
regularity.  Further component sectors are collected in the
Supplemental Material.

\section*{Discussion and outlook}
\label{sec:discussion}

We have shown that fermionic hidden zeros obey the same rectangular
kinematic geometry as their bosonic counterparts, with particle species
entering only through a finite set of local bridge data.  The result holds
at arbitrary multiplicity and for general rectangular loci, including all
multi-pinch sectors, and extends naturally to graviton--gravitino sectors
and mixed gauge--gravity states.  In particular, the fermion--fermion
sector introduces no proliferation of higher spinorial bridge structures:
the hidden-zero bridge remains entirely governed by the vector current.

Several nearby structures deserve further investigation.  It would be
interesting to understand how the species-dependent bridges fit into the
2-split framework \cite{Cao:2024gln,Cao:2024qpp,Azevedo:2025vxo}, and
whether they admit a natural description in scaffolding and surface
formulations \cite{Arkani-Hamed:2023jry,De:2024wsy,Cao:2025lzv,
Chang:2025cqe}.  At finite $\alpha'$, open-superstring amplitudes are
organized in terms of SYM data and admit direct connections to
field-theory and CHY representations
\cite{Mafra:2011nv,He:2018pol}; related developments include
factorization near hidden-zero loci
\cite{Zhang:NewFactorizationToAppear} and a universal finite-$\alpha'$
worldsheet description of hidden-zero and 2-split structures
\cite{He:FiniteAlphaPrimeToAppear}.  It would also be interesting to
understand how the fermionic bridge structure is related to soft behavior
and BCJ amplitude relations
\cite{Weinberg:1965nx,Bern:2008qj,Guevara:2024nxd,Bartsch:2024amu},
and how it behaves under double-copy constructions
\cite{Bern:2010ue,Bartsch:2024amu,Li:2024qfp}.

A complementary goal is a purely field-theoretic derivation.  Bootstrap,
locality, and uniqueness approaches
\cite{Rodina:2024yfc,Zhou:2026isc,Rodina:2026yaq}, direct Feynman-diagram
analyses \cite{Zhou:2024ddy,Zhou:2026fkg}, and amplitude-expansion or
recursive constructions
\cite{Huang:2025blb,Feng:2025dci,Jones:2025rbv,Feng:2025ofq}
have already provided complementary descriptions of hidden zeros and
splittings.  Beyond this, the extensions to Dirac and fundamental
fermions suggest QCD-like applications, with colored-Yukawa surface
formulations providing a natural neighboring framework
\cite{De:2024wsy}; related hidden-zero structures have also appeared in
cosmological wavefunctions \cite{Li:2026gns,De:2025bmf}, loop integrands
\cite{Backus:2025hpn}, and geometric curve and surface formulations
\cite{Arkani-Hamed:2023lbd,Arkani-Hamed:2023mvg,Arkani-Hamed:2024nhp,
Arkani-Hamed:2024yvu,Arkani-Hamed:2024tzl}.

\section*{Acknowledgements}

We thank Qu Cao and Fanky Zhu for helpful discussions.  We are particularly
grateful to Song He for many stimulating discussions and for related
collaboration.  This work is supported by the National Natural Science
Foundation of China under Grant No.~12405086.

\bibliographystyle{physics}

\bibliography{Refs}

%\end{document}

\widetext

\clearpage

\onecolumngrid

\begin{center}
{\large\bfseries SUPPLEMENTARY MATERIAL}
\end{center}

\vspace{1.5em}

%\twocolumngrid

\section{Conventions and representative low-point checks}
\label{app:conventions-lowpoint}
\label{app:conventions}
\label{app:low-point-examples}

We collect here the spinor conventions used throughout the paper and a
small set of representative component checks of the species-dependent
hidden-zero conditions.  The examples are chosen to illustrate the
distinct local bridge structures without exhaustively enumerating all
rectangular loci at low multiplicity.

\subsection{Spinor conventions}

We use a mostly-minus metric and an all-outgoing momentum convention, and
define
\begin{equation}
\slashed v:=v_\mu\gamma^\mu .
\label{eq:app-slash-convention}
\end{equation}
The gamma matrices satisfy
\begin{equation}
\{\gamma^\mu,\gamma^\nu\}
=
2\eta^{\mu\nu},
\qquad
\gamma^{\mu\nu}
=
\frac12[\gamma^\mu,\gamma^\nu].
\label{eq:app-clifford}
\end{equation}
An on-shell massless fermion wavefunction obeys
\begin{equation}
k_i^2=0,
\qquad
\slashed k_i\chi_i=0 .
\label{eq:app-dirac}
\end{equation}

Spinor contractions are written with the charge-conjugation structure
implicitly understood.  In particular, we define
\begin{equation}
J_{ij}^{\mu}
:=
\chi_i\gamma^\mu\chi_j
\equiv
\chi_i^\alpha
(C\gamma^\mu)_{\alpha\beta}
\chi_j^\beta .
\label{eq:app-vector-bilinear}
\end{equation}
External polarization spinors are treated as commuting c-number
wavefunctions.  With the charge-conjugation conventions used here,
including the implicit symplectic contraction in $D=6$, the effective
vector bilinear is symmetric,
\begin{equation}
J_{ij}^{\mu}=J_{ji}^{\mu}.
\label{eq:app-current-exchange}
\end{equation}
Both left- and right-acting on-shell Dirac equations are understood in the
spinor manipulations below.

We consider a single adjoint fermion species and use the corresponding
Majorana-type massless representations: Majorana spinors in $D=4$,
symplectic Majorana--Weyl spinors in $D=6$, and Majorana--Weyl spinors in
$D=10$.  The appropriate charge-conjugation, reality, chirality, and, in
six dimensions, symplectic-index contractions are understood in
Eq.~\eqref{eq:app-vector-bilinear}.  The symplectic index in $D=6$ is part
of the reality structure and is not treated as an independent flavor
label.

A common feature of these minimal-SYM spinor representations is the
three-spinor Fierz identity.  In our conventions it may be written as
\begin{equation}
(C\gamma^\mu)_{\alpha(\beta}
(C\gamma_\mu)_{\gamma\delta)}
=
0,
\qquad
D=4,6,10 .
\label{eq:app-three-spinor-fierz}
\end{equation}
Contracting with four external wavefunctions gives
\begin{equation}
J_{12}\cdot J_{34}
+
J_{13}\cdot J_{42}
+
J_{14}\cdot J_{23}
=
0 .
\label{eq:app-vector-current-fierz}
\end{equation}
Different dimension-specific conventions may introduce a common nonzero
normalization of the bilinear; we absorb it into the definition
\eqref{eq:app-vector-bilinear}.  In particular, the bridge condition
\begin{equation}
J_{ij}^{\mu}=0
\label{eq:app-vector-current-zero}
\end{equation}
is insensitive to such an overall normalization.

For a gluon of momentum $k_i$ and polarization $\epsilon_i^\mu$ we use
\begin{equation}
k_i\cdot\epsilon_i=0,
\qquad
f_i^{\mu\nu}
=
k_i^\mu\epsilon_i^\nu-k_i^\nu\epsilon_i^\mu ,
\label{eq:app-gluon-conventions}
\end{equation}
together with
\begin{equation}
\slashed f_i
=
\frac14 f_{i,\mu\nu}\gamma^{\mu\nu}.
\label{eq:app-fslash}
\end{equation}
In the component amplitudes below we suppress common couplings, phases,
and convention-dependent overall numerical factors, but never relative
graph weights.  Factors of two arising from
$(k_i+k_j)^2=2s_{ij}$ are likewise absorbed into the overall
normalization of the displayed amplitudes.

\subsection{Representative low-point checks}
\label{app:representative-lowpoint}

We now give a few component-field realizations of the four local bridge
types.  At four points the two inequivalent rectangular geometries for
the cyclic ordering $(1,2,3,4)$ may be represented by
\begin{equation}
(a,b)=(4,2):
\quad
L=\{1\},
\quad
R=\{3\},
\qquad
(a,b)=(1,3):
\quad
L=\{2\},
\quad
R=\{4\}.
\label{eq:app-fourpoint-rectangles}
\end{equation}
Thus the active cross pair is respectively $(1,3)$ or $(2,4)$.
The examples below show how different particle assignments decorate
these same two geometries.

\medskip
\noindent
\underline{\emph{$ffgg$: a mixed bridge.}}

Consider
\begin{equation}
A_4(1_f,2_f,3_g,4_g).
\label{eq:app-fourpoint-ffgg}
\end{equation}
For $(a,b)=(4,2)$ the active pair is $(1_f,3_g)$, and the hidden-zero
conditions are
\begin{equation}
s_{13}=0,
\qquad
k_1\cdot\epsilon_3=0,
\qquad
\chi_1\slashed f_3=0 .
\label{eq:app-fourpoint-ffgg-zero}
\end{equation}
The two planar diagrams give
\begin{align}
A_4^{(g)}
&=
\frac{1}{s_{12}}\,
J_{12}^{\mu}
V_{\mu\nu\rho}(k_1+k_2,k_3,k_4)
\epsilon_3^\nu\epsilon_4^\rho ,
\nonumber\\
A_4^{(f)}
&=
-\frac{1}{s_{23}}\,
\chi_1\slashed\epsilon_4
(\slashed k_2+\slashed k_3)
\slashed\epsilon_3\chi_2 ,
\label{eq:app-fourpoint-ffgg-feynman}
\end{align}
where
\begin{equation}
V_{\mu\nu\rho}(p,q,r)
=
\eta_{\mu\nu}(p-q)_\rho
+
\eta_{\nu\rho}(q-r)_\mu
+
\eta_{\rho\mu}(r-p)_\nu ,
\qquad
p+q+r=0 .
\label{eq:app-three-gluon-vertex}
\end{equation}
On $s_{13}=0$, momentum conservation gives $s_{23}=-s_{12}$, and the
complete amplitude reduces to
\begin{equation}
\left.
A_4(1_f,2_f,3_g,4_g)
\right|_{s_{13}=0}
=
\frac{1}{s_{12}}
\left[
\chi_1\slashed f_3\slashed\epsilon_4\chi_2
-
(k_1\cdot\epsilon_3)
\chi_1\slashed\epsilon_4\chi_2
\right].
\label{eq:app-fourpoint-ffgg-check}
\end{equation}
The two terms are annihilated separately by the two components of the
$f_1g_3$ bridge, and hence
\begin{equation}
\left.
A_4(1_f,2_f,3_g,4_g)
\right|_{
s_{13}=0,\,
k_1\cdot\epsilon_3=0,\,
\chi_1\slashed f_3=0
}
=0 .
\end{equation}

\medskip
\noindent
\underline{\emph{$fgfg$: different bridges from the same ordering.}}

Next consider
\begin{equation}
A_4(1_f,2_g,3_f,4_g).
\label{eq:app-fourpoint-fgfg}
\end{equation}
For $(a,b)=(4,2)$ the active pair is $(1_f,3_f)$, giving
\begin{equation}
s_{13}=0,
\qquad
J_{13}^{\mu}=0 .
\label{eq:app-fourpoint-fgfg-zero}
\end{equation}
The two planar fermion-exchange diagrams give
\begin{equation}
A_4
=
\frac{
\chi_1\slashed\epsilon_2
(\slashed k_1+\slashed k_2)
\slashed\epsilon_4\chi_3
}{s_{12}}
+
\frac{
\chi_1\slashed\epsilon_4
(\slashed k_1+\slashed k_4)
\slashed\epsilon_2\chi_3
}{s_{14}} .
\label{eq:app-fourpoint-fgfg-feynman}
\end{equation}
On $s_{13}=0$, with $s_{14}=-s_{12}$,
\begin{equation}
\left.
A_4(1_f,2_g,3_f,4_g)
\right|_{s_{13}=0}
=
\frac{1}{s_{12}}\,
J_{13}^{\mu}
\left[
2(k_1+k_3)\cdot\epsilon_2\,\epsilon_{4\mu}
-
2(k_1+k_3)\cdot\epsilon_4\,\epsilon_{2\mu}
-
(\epsilon_2\cdot\epsilon_4)(k_2-k_4)_\mu
\right].
\label{eq:app-fourpoint-fgfg-check}
\end{equation}
Thus
\begin{equation}
\left.
A_4(1_f,2_g,3_f,4_g)
\right|_{
s_{13}=0,\,
J_{13}^{\mu}=0
}
=0 .
\end{equation}

For the other rectangular geometry, $(a,b)=(1,3)$, the active pair is
instead $(2_g,4_g)$, and the same amplitude possesses the distinct
gluonic hidden zero
\begin{equation}
s_{24}=0,
\qquad
\epsilon_2\cdot\epsilon_4=0,
\qquad
\epsilon_2\cdot k_4=0,
\qquad
k_2\cdot\epsilon_4=0 .
\label{eq:app-fourpoint-fgfg-gg-zero}
\end{equation}
The alternating ordering therefore makes explicit that different
rectangles of the same physical amplitude can expose different local
bridge types.

\medskip
\noindent
\underline{\emph{$ffff$: Fierz routing to the active bridge.}}

Finally consider
\begin{equation}
A_4(1_f,2_f,3_f,4_f).
\label{eq:app-fourpoint-ffff}
\end{equation}
For $(a,b)=(4,2)$ the active pair is $(1_f,3_f)$:
\begin{equation}
s_{13}=0,
\qquad
J_{13}^{\mu}=0 .
\label{eq:app-fourpoint-ffff-zero}
\end{equation}
The two planar gluon-exchange diagrams give
\begin{equation}
A_4
=
\frac{J_{12}\cdot J_{34}}{s_{12}}
-
\frac{J_{23}\cdot J_{41}}{s_{23}} .
\label{eq:app-fourpoint-ffff-feynman}
\end{equation}
On $s_{13}=0$, with $s_{23}=-s_{12}$,
\begin{equation}
\left.
A_4
\right|_{s_{13}=0}
=
\frac{1}{s_{12}}
\left(
J_{12}\cdot J_{34}
+
J_{23}\cdot J_{41}
\right).
\label{eq:app-fourpoint-ffff-before-fierz}
\end{equation}
Using Eq.~\eqref{eq:app-vector-current-fierz},
\begin{equation}
\left.
A_4(1_f,2_f,3_f,4_f)
\right|_{s_{13}=0}
=
-\frac{1}{s_{12}}\,
J_{13}\cdot J_{24},
\label{eq:app-fourpoint-ffff-check}
\end{equation}
and therefore
\begin{equation}
\left.
A_4(1_f,2_f,3_f,4_f)
\right|_{
s_{13}=0,\,
J_{13}^{\mu}=0
}
=0 ,
\qquad
D=4,6,10 .
\end{equation}
This example illustrates explicitly how the Fierz identity reroutes the
planar fermion-line pairings into the vector-current bridge selected by
the rectangular geometry.

\medskip
\noindent
\underline{\emph{A six-point mixed example.}}

A single higher-point example is useful for showing that the four bridge
types are not independent constructions.  Consider
\begin{equation}
A_6(1_f,2_g,3_g,4_f,5_g,6_g)
\label{eq:app-six-point-amplitude}
\end{equation}
with distinguished legs
\begin{equation}
(a,b)=(6,3).
\end{equation}
Then
\begin{equation}
L(6,3)=\{1_f,2_g\},
\qquad
R(3,6)=\{4_f,5_g\},
\end{equation}
so that
\begin{equation}
L(6,3)\times R(3,6)
=
\left\{
(1_f,4_f),\,
(1_f,5_g),\,
(2_g,4_f),\,
(2_g,5_g)
\right\}.
\label{eq:app-six-point-cross-pairs}
\end{equation}
The four cross pairs realize respectively the
$ff$, $fg$, $gf$, and $gg$ bridges.

The rectangular kinematics is
\begin{equation}
s_{14}=s_{15}=s_{24}=s_{25}=0 ,
\label{eq:app-six-point-kinematics}
\end{equation}
while the corresponding bridge conditions are
\begin{align}
&\chi_1\gamma^\mu\chi_4=0,
\nonumber\\
&k_1\cdot\epsilon_5=0,
\qquad
\chi_1\slashed f_5=0,
\nonumber\\
&\epsilon_2\cdot k_4=0,
\qquad
\slashed f_2\chi_4=0,
\nonumber\\
&\epsilon_2\cdot\epsilon_5=0,
\qquad
\epsilon_2\cdot k_5=0,
\qquad
k_2\cdot\epsilon_5=0 .
\label{eq:app-six-point-bridges}
\end{align}
The general theorem therefore gives
\begin{equation}
\left.
A_6(1_f,2_g,3_g,4_f,5_g,6_g)
\right|_{
\substack{
s_{14}=s_{15}=s_{24}=s_{25}=0\\
x_{14}=x_{15}=x_{24}=x_{25}=0
}
}
=0 .
\label{eq:app-six-point-check}
\end{equation}

Thus already at six points all four species-dependent bridges occur
simultaneously as local decorations of a single rectangular geometry.
No additional bridge structure is required.

\section{Beyond Majorana fermions}
\label{app:beyond-majorana}

The Majorana-type reality condition used in the main text removes
fermion-line orientation as an independent datum: every external fermionic
state is denoted by $f$, and any fermion--fermion cross pair may participate
in the vector fusion $ff\to g$.  We briefly describe how this structure is
modified when fermion number and color-flow orientation are restored.  The
single-line sector retains the same local Lorentz structures, while amplitudes
with several Dirac lines are naturally organized into relations among
charge-resolved components.

\subsection{Single-fermion-line extensions}
\label{app:single-fermion-line}

Consider first an amplitude containing one fermion, one antifermion, and an
arbitrary number of gluons.  For adjoint Dirac matter the two fermionic legs
may occupy arbitrary positions in the color ordering,
\begin{equation}
A_n(\ldots,i_f,\ldots,j_{\bar f},\ldots),
\qquad
A_n(\ldots,i_{\bar f},\ldots,j_f,\ldots).
\label{eq:app-single-dirac-line}
\end{equation}
In an all-outgoing convention we associate a column spinor $u_i$ with an
external fermion and a row spinor $\bar v_i$ with an external antifermion,
\begin{equation}
f_i:\ u_i,
\qquad
\bar f_i:\ \bar v_i,
\qquad
\slashed k_i u_i=0,
\qquad
\bar v_i\slashed k_i=0 .
\label{eq:app-dirac-wavefunctions}
\end{equation}

The fermion--antifermion fusion is the oriented counterpart of the Majorana
channel,
\begin{equation}
f_i\bar f_j\longrightarrow g,
\end{equation}
with vector-current coefficient
\begin{equation}
J_{i\bar j}^{\mu}
:=
\bar v_j\gamma^\mu u_i .
\label{eq:app-dirac-vector-current}
\end{equation}
Hence the corresponding bridge is
\begin{equation}
f_i\bar f_j:
\qquad
\bar v_j\gamma^\mu u_i=0 ,
\label{eq:app-dirac-fbarf-bridge}
\end{equation}
with the reversed orientation
$\bar f_i f_j$ carrying
$\bar v_i\gamma^\mu u_j=0$.

The mixed fermion--gluon bridges retain the same local Lorentz structures as
in the Majorana theory.  Collecting all oriented possibilities,
\begin{equation}
\begin{array}{c|l}
\text{cross pair} & x_{ij} \\ \hline
f_i\bar f_j
&
\left\{\bar v_j\gamma^\mu u_i\right\}
\\[1mm]
\bar f_i f_j
&
\left\{\bar v_i\gamma^\mu u_j\right\}
\\[1mm]
f_i g_j
&
\left\{k_i\cdot\epsilon_j,\,
\slashed f_j u_i\right\}
\\[1mm]
g_i f_j
&
\left\{\epsilon_i\cdot k_j,\,
\slashed f_i u_j\right\}
\\[1mm]
\bar f_i g_j
&
\left\{k_i\cdot\epsilon_j,\,
\bar v_i\slashed f_j\right\}
\\[1mm]
g_i\bar f_j
&
\left\{\epsilon_i\cdot k_j,\,
\bar v_j\slashed f_i\right\}
\\[1mm]
g_i g_j
&
\left\{
\epsilon_i\cdot\epsilon_j,\,
\epsilon_i\cdot k_j,\,
k_i\cdot\epsilon_j
\right\}.
\end{array}
\label{eq:app-single-line-dirac-bridges}
\end{equation}
Thus orientation is added to the bridge data, but no new primitive Lorentz
structure appears.  The local fusion algebra is simply
\begin{equation}
f\bar f\longrightarrow g,
\qquad
fg/gf\longrightarrow f,
\qquad
gg\longrightarrow g .
\label{eq:app-single-line-fusions}
\end{equation}

A simple four-point example is
\begin{equation}
A_4(1_f,2_g,3_{\bar f},4_g).
\label{eq:app-dirac-fourpoint-fgfbarg}
\end{equation}
For the rectangle exposing the pair $(1_f,3_{\bar f})$,
\begin{equation}
s_{13}=0,
\qquad
\bar v_3\gamma^\mu u_1=0
\label{eq:app-dirac-fourpoint-fbarf-zero}
\end{equation}
gives the oriented fermion--antifermion hidden zero.  The complementary
rectangle exposes instead the gluonic pair $(2_g,4_g)$ and therefore carries
the usual $gg$ bridge.  This already illustrates that the oriented Dirac
description changes the species labels of the bridges without changing their
local structure.

\medskip
\noindent\emph{Fundamental fermions.}

For a single fermion line, the passage from adjoint to fundamental matter is
particularly simple.  For a fixed-endpoint ordering,
\begin{equation}
A_n^{\rm adj}
\bigl(1_f,\alpha,n_{\bar f}\bigr)
=
A_n^{\rm prim}
\bigl(1_q,\alpha,n_{\bar q}\bigr)
\label{eq:app-adjoint-fundamental-kinematic-equality}
\end{equation}
up to the common normalization conventions used throughout this work.  The
two amplitudes have the same planar propagators and color-stripped vertices;
only their global color organization differs.  Consequently, every
open-compatible adjoint-Dirac hidden zero is inherited directly by the
corresponding fundamental primitive under
\begin{equation}
f\longrightarrow q,
\qquad
\bar f\longrightarrow\bar q .
\label{eq:app-fundamental-replacement}
\end{equation}

A general adjoint one-line ordering may instead place gluons on both sides of
the fermion pair,
\begin{equation}
A_n^{\rm adj}
\bigl(1_f,\alpha,n_{\bar f},\beta\bigr).
\end{equation}
It can be rewritten in the fixed-endpoint fundamental basis as
\begin{equation}
A_n^{\rm adj}
\bigl(1_f,\alpha,n_{\bar f},\beta\bigr)
=
(-1)^{|\beta|}
\sum_{\sigma\in\alpha\shuffle\beta^T}
A_n^{\rm prim}
\bigl(1_q,\sigma,n_{\bar q}\bigr),
\label{eq:app-general-adjoint-fundamental-relation}
\end{equation}
where $\beta^T$ denotes the reversed word.  Hence an adjoint hidden zero that
is not compatible with a single open primitive becomes instead the linear
relation
\begin{equation}
\left.
\sum_{\sigma\in\alpha\shuffle\beta^T}
A_n^{\rm prim}
\bigl(1_q,\sigma,n_{\bar q}\bigr)
\right|_{H_{L|R}}
=0 .
\label{eq:app-general-fundamental-primitive-relation}
\end{equation}
Thus changing the color representation reorganizes, rather than removes, the
single-line hidden-zero information.

\medskip
\noindent\emph{Massive one-line matter.}

The same one-line reasoning admits a simple massive extension.  Dimensional
reduction from a higher-dimensional massless momentum shows that a massive
fermion--gluon cross pair retains both the massless rectangular condition
\begin{equation}
k_f\cdot k_g=0
\end{equation}
and the same mixed bridge.  For an equal-mass fermion--antifermion pair with
opposite internal momenta, however, the rectangular kinematics is shifted to
\begin{equation}
s_{f\bar f}=-m^2,
\qquad\Longleftrightarrow\qquad
(k_f+k_{\bar f})^2=0 ,
\label{eq:app-massive-ffbar-kinematics}
\end{equation}
while the vector-current bridge remains unchanged,
\begin{equation}
\bar v_{\bar f}\gamma^\mu u_f=0 .
\end{equation}
The corresponding open-compatible fundamental primitives inherit the same
massive zeros.  We do not pursue the massive sector further here.

\subsection{Multiple fermion lines and Dirac relations}
\label{app:adjoint-dirac-relations}

With several fermion lines, fermion-number orientation becomes genuine global
data.  A real adjoint fermion amplitude therefore need not map to the
vanishing of each charge-resolved Dirac component separately.  Instead, the
Majorana hidden zero becomes a linear relation among such components.

To make this explicit, combine two degenerate real adjoint fermions into a
Dirac field,
\begin{equation}
\Psi
=
\frac{1}{\sqrt2}
\left(
\chi^1+i\chi^2
\right),
\qquad
\chi^1
=
\frac{1}{\sqrt2}
\left(
\Psi+\Psi^c
\right).
\label{eq:app-majorana-dirac-basis}
\end{equation}
For $2r$ fixed fermionic slots
$F=\{i_1,\ldots,i_{2r}\}$, multilinearity gives
\begin{equation}
A_n\bigl(\{\chi^1_{i_a}\}_{a=1}^{2r};\,g_{\bar F}\bigr)
=
2^{-r}
\sum_{\substack{
Q_{i_a}\in\{f,\bar f\}\\
N_f=N_{\bar f}=r
}}
A_n\bigl(\{Q_{i_a}\}_{a=1}^{2r};\,g_{\bar F}\bigr).
\label{eq:app-general-dirac-basis-expansion}
\end{equation}
The sum runs over the
\begin{equation}
\binom{2r}{r}
\end{equation}
charge-neutral assignments of the fixed fermionic slots.

The Majorana vector bridge is correspondingly resolved into the two oriented
Dirac currents,
\begin{equation}
J_{ij}^{\mu}(\chi^1)
=
\frac12
\left(
\bar v_j\gamma^\mu u_i
+
\bar v_i\gamma^\mu u_j
\right).
\label{eq:app-majorana-current-dirac-decomposition}
\end{equation}
Consequently every Majorana hidden zero induces the charge-sector relation
\begin{equation}
\left.
\sum_{\substack{
Q_{i_a}\in\{f,\bar f\}\\
N_f=N_{\bar f}=r
}}
A_n\bigl(\{Q_{i_a}\}_{a=1}^{2r};\,g_{\bar F}\bigr)
\right|_{H_{L|R}^{(\chi^1)}}
=0 ,
\label{eq:app-general-dirac-amplitude-relation}
\end{equation}
where the Majorana bridge conditions on
$H_{L|R}^{(\chi^1)}$ are expressed through the oriented currents according to
Eq.~\eqref{eq:app-majorana-current-dirac-decomposition}.

Thus the hidden-zero structure is basis covariant: in the real Majorana basis
it appears as the vanishing of an individual component, whereas in the Dirac
basis the same statement becomes a relation among charge-resolved
amplitudes.  A systematic treatment of several fundamental fermion lines,
where flavor and open color-flow routing introduce additional global data,
is left for future work.

\section{Six- and eight-fermion component checks in ten dimensions}
\label{app:higher-fermion-checks}

We summarize here independent component-field checks of the fermionic
hidden-zero theorem at six and eight external fermions.  With computer-algebra assistance, two independent computational
implementations were developed directly from the ten-dimensional SYM
action: a color-ordered Feynman-diagram expansion and a Berends--Giele
reconstruction.  The final reductions were performed with exact symbolic
algebra and do not invoke the CHY or RNS representations used in the main
text.  They therefore
provide complementary field-theoretic checks of the bridge conditions at
substantially higher fermion multiplicity.

The six-fermion amplitude already requires nontrivial ten-dimensional
Fierz routing among different fermion-line pairings.  At eight fermions
the calculation becomes considerably more stringent: genuine rank-five
Clifford structures occur before the full bridge relations are imposed,
so the vanishing cannot be understood as a term-by-term reduction to
vector currents.  We collect below the main ingredients and exact results
of these two calculations. 

\subsection{Six fermions}
\label{app:six-fermion-check}

We first consider
\begin{equation}
A_6(1_f,2_f,3_f,4_f,5_f,6_f)
\end{equation}
in ten-dimensional SYM.  The direct color-ordered expansion contains
fourteen planar tree graphs.  Twelve contain four $ffg$ vertices and an
internal fermion propagator, while two additional ``Mercedes'' graphs
contain three $ffg$ vertices and one $ggg$ vertex.  The independent
Berends--Giele reconstruction reproduces the same fourteen graphs,
including their propagators and relative signs.

The reduction uses the massless Dirac equations, momentum conservation,
ordinary Clifford algebra, and the ten-dimensional chiral three-spinor
identity
\begin{equation}
(C\gamma^\mu)_{\alpha(\beta}
(C\gamma_\mu)_{\gamma\delta)}
=0 .
\label{eq:app-six-fierz}
\end{equation}
In open-chain form this gives
\begin{equation}
J_{ab}^{\mu}
\bigl(\chi_c\gamma_\mu X\chi_d\bigr)
+
J_{ac}^{\mu}
\bigl(\chi_b\gamma_\mu X\chi_d\bigr)
+
\bigl(\chi_a\gamma_\mu X\chi_d\bigr)
J_{bc}^{\mu}
=0 ,
\label{eq:app-six-open-fierz}
\end{equation}
where $J_{ij}^{\mu}$ is defined in
Eq.~\eqref{eq:app-vector-bilinear} and $X$ is an even gamma word.
This identity is needed to route three-form open-chain structures among
different fermion-line pairings; a scalar four-spinor rearrangement
alone is not sufficient.

We tested two inequivalent rectangular loci.  In this subsection we
abbreviate the canonical hidden-zero locus by
\begin{equation}
H_m:=H_{L_m|R_m},
\label{eq:app-Hm}
\end{equation}
and distinguish it from the corresponding Mandelstam sublocus $h_m$.
For $m=1$, define
\begin{equation}
h_1
:=
\left\{
s_{13}=s_{14}=s_{15}=0
\right\},
\qquad
\mathcal J_1
:=
\left\langle
J_{13}^{\mu},
J_{14}^{\mu},
J_{15}^{\mu}
\right\rangle ,
\label{eq:app-six-h1}
\end{equation}
where $\mathcal J_1$ is the ideal generated by the vector-current
bridges.  Thus the complete locus $H_1$ is obtained from $h_1$ by
additionally imposing the generators of $\mathcal J_1$.

On the Mandelstam sublocus, the complete amplitude reduces to
\begin{equation}
\left.
A_6
\right|_{h_1}
=
J_{13}^{\mu}B_{\mu}^{(13)}
+
J_{14}^{\mu}B_{\mu}^{(14)}
+
J_{15}^{\mu}B_{\mu}^{(15)},
\label{eq:app-six-m1-reduction}
\end{equation}
where the coefficients $B_{\mu}^{(ij)}$ are rational functions whose
denominators contain only ordinary planar propagators.  Away from these
poles,
\begin{equation}
\left.
A_6
\right|_{h_1}
\in
\mathcal J_1
\qquad\Longrightarrow\qquad
\left.
A_6
\right|_{H_1}
=0 .
\label{eq:app-six-m1-result}
\end{equation}

For $m=2$, similarly define
\begin{equation}
h_2
:=
\left\{
s_{14}=s_{15}=s_{24}=s_{25}=0
\right\},
\qquad
\mathcal J_2
:=
\left\langle
J_{14}^{\mu},
J_{15}^{\mu},
J_{24}^{\mu},
J_{25}^{\mu}
\right\rangle .
\label{eq:app-six-h2}
\end{equation}
The complete reduction gives
\begin{equation}
\left.
A_6
\right|_{h_2}
=
\sum_{(i,j)\in
\{(1,4),(1,5),(2,4),(2,5)\}}
J_{ij}^{\mu}B_{\mu}^{(ij)},
\label{eq:app-six-m2-reduction}
\end{equation}
and hence, away from ordinary planar poles,
\begin{equation}
\left.
A_6
\right|_{h_2}
\in
\mathcal J_2
\qquad\Longrightarrow\qquad
\left.
A_6
\right|_{H_2}
=0 .
\label{eq:app-six-m2-result}
\end{equation}
This reduction includes the two row/column-disjoint bridge configurations
\begin{equation}
\{(1,4),(2,5)\},
\qquad
\{(1,5),(2,4)\},
\end{equation}
which must be kept distinct during the Fierz routing.

The cancellation is not graph by graph.  Three-form structures from
different fermion-line pairings mix under
Eq.~\eqref{eq:app-six-open-fierz}, while the two Mercedes graphs supply
essential vector contributions.  After the complete fourteen-graph sum,
however, no structure survives outside the vector-current bridge ideal.
The six-fermion calculation therefore confirms directly that the same
local bridge $J_{ij}^{\mu}=0$ controls both the $m=1$ and $m=2$
rectangular configurations.

\subsection{Eight fermions}
\label{app:eight-fermion-check}

We next consider
\begin{equation}
A_8(1_f,2_f,3_f,4_f,5_f,6_f,7_f,8_f)
\end{equation}
for the $m=3$ rectangle
\begin{equation}
L_3=\{1,2,3\},
\qquad
R_3=\{5,6,7\}.
\end{equation}
Its Mandelstam sublocus is
\begin{equation}
h_3
:=
\left\{
s_{ij}=0
\ \middle|\
i=1,2,3,\quad j=5,6,7
\right\},
\label{eq:app-eight-h3}
\end{equation}
and the corresponding vector-current bridge ideal is
\begin{equation}
\mathcal J_3
:=
\left\langle
J_{ij}^{\mu}
\ \middle|\
i=1,2,3,\quad j=5,6,7
\right\rangle .
\label{eq:app-eight-current-ideal}
\end{equation}
Thus the test involves nine cross-pair Mandelstam conditions and nine
vector-current bridges, with the full hidden-zero locus $H_3$ obtained
by imposing both.

The complete color-ordered component expansion contains 134 planar
contributions,
\begin{equation}
\begin{array}{c|c}
\text{vertex content} & \text{number of graphs} \\ \hline
6\,ffg & 96 \\
5\,ffg+1\,ggg & 32 \\
4\,ffg+2\,ggg & 4 \\
4\,ffg+1\,gggg & 2
\end{array}
\label{eq:app-eight-topologies}
\end{equation}
and the same result is reproduced independently by the component
Berends--Giele recursion.  Both the cubic gluon self-interaction and the
quartic YM vertex are required.

The eight-fermion reduction is qualitatively more involved than the
six-fermion case.  A fermion line containing two internal fermion
propagators generates five-gamma chains of the form
\begin{equation}
\gamma^\mu
\slashed P\,
\gamma^\nu
\slashed Q\,
\gamma^\rho ,
\end{equation}
whose Clifford decomposition contains a genuine rank-five component.
For same-chirality ten-dimensional spinors the relevant sectors are
schematically
\begin{equation}
S_+\otimes S_+
\supset
V\oplus\Lambda^3V\oplus\Lambda^5_+V .
\label{eq:app-eight-clifford-sectors}
\end{equation}
Thus $J_{ij}^{\mu}=0$ does not separately force an individual
five-form bilinear $\chi_i\gamma^{[5]}\chi_j$ to vanish.  Indeed,
nonzero rank-five coefficients are present before the complete bridge
relations are imposed.  The eight-fermion zero is therefore a property
of the full amplitude rather than a graph-by-graph or
Clifford-component-by-component cancellation.

For the exact algebraic check, let $\mathcal P$ be the polynomial ring
over $\mathbb Q(i)$ in the momentum and positive-chirality spinor
coordinates.  We define the full bridge--Dirac--momentum ideal
\begin{equation}
I_B
=
\left\langle
k_i^2,\,
\sum_{i=1}^8 k_i,\,
\slashed k_i\chi_i,\,
s_{ij},\,
J_{ij}^{\mu}
\right\rangle ,
\qquad
i=1,\ldots,8,
\quad
(i,j)\in L_3\times R_3 ,
\label{eq:app-eight-bridge-ideal}
\end{equation}
with all vector and spinor equations understood componentwise.  The
cross-pair generators $s_{ij}$ and $J_{ij}^{\mu}$ encode precisely the
conditions defining $H_3$, while the remaining generators impose the
on-shell Dirac equations and momentum conservation.

Let
\begin{equation}
S
=
\prod_{\alpha}P_\alpha
\label{eq:app-eight-planar-product}
\end{equation}
be the product of all ordinary planar propagator polynomials appearing
in the amplitude, and let $N_8$ denote the complete numerator obtained
after clearing these propagators.  The exact calculation gives
\begin{equation}
N_8
\in
\sqrt{\,I_B:S^\infty\,}.
\label{eq:app-eight-radical-membership}
\end{equation}
Equivalently, the cleared numerator vanishes on the complex variety
defined by the bridge, Dirac, and momentum constraints after excluding
components supported entirely on ordinary planar factorization poles.
In particular,
\begin{equation}
\left.
A_8(1_f,\ldots,8_f)
\right|_{H_3}
=0
\qquad
\text{away from ordinary planar poles}.
\label{eq:app-eight-hidden-zero}
\end{equation}

To establish Eq.~\eqref{eq:app-eight-radical-membership}, the exact
calculation decomposes the localized bridge variety into the relevant
spinor-geometric components.  On the dominant component, the
vector-current bridge equations enforce a common-null structure, which
removes the native five-form contribution and reduces the remaining
numerator to lower-point identities.  The pure-spinor component vanishes
separately, while the remaining mixed bridge components are checked by
exact characteristic-zero Clifford and numerator reductions.  Together,
these cases cover the localized bridge variety and establish the radical
membership above.

A complete proof of this localized radical-membership statement, together
with exact computational certificates and reproducibility instructions, is
provided in the ancillary material.

\section{Worldsheet details}
\label{app:worldsheet-details}

We collect here the technical worldsheet ingredients underlying the
main-text worldsheet proof.  The role of the worldsheet analysis is
local: for each pinched cross pair, the behavior of the RNS
half-integrand is determined by the OPE of the two colliding external
states.  We first identify the local fusion channels and show that the
species-dependent bridge data of Eq.~\eqref{eq:species-bridge} remove
all contributions with negative total scaling on the regulated
rectangular degeneration.  We then explain why picture changing
introduces no additional primitive divergent fermionic structures at
arbitrary Ramond multiplicity.  Finally, we show that the same regulated
$O(1)$ behavior holds jointly on a general multi-pinch sector.

The formulas below are written in the ten-dimensional RNS ambitwistor
language.  Locally, the NS and Ramond vertices, spin fields, and
picture-changing algebra closely resemble their conventional RNS
counterparts, but the underlying worldsheet model is the ambitwistor
string rather than the finite-$\alpha'$ superstring.  In $D=4,6$, the
corresponding minimal-spinor OPEs have the same fermionic
principal-part structure: the leading fermion--fermion singularity is
again the vector channel, while higher-rank channels are regular at the
collision.  They therefore reproduce the same spacetime bridge data
used in the main text.

\subsection{RNS vertices and local OPEs}
\label{app:rns-local-opes}

For the NS and Ramond external states we use the standard RNS vertex
operators, with the gluon--gluino disk-amplitude conventions and
supersymmetry relations reviewed for example in
Ref.~\cite{Stieberger:2007jv},
\begin{align}
V_g^{(-1)}(\epsilon,k)
&=
\epsilon\cdot\psi\,
e^{-\phi}e^{ik\cdot X},
\nonumber\\
V_g^{(0)}(\epsilon,k)
&=
\left(
\epsilon\cdot P
+\frac12 f_{\mu\nu}\psi^\mu\psi^\nu
\right)e^{ik\cdot X},
\nonumber\\
V_f^{(-1/2)}(\chi,k)
&=
2^{-1/4}\chi^\alpha S_\alpha
e^{-\phi/2}e^{ik\cdot X}.
\label{eq:app-rns-vertices}
\end{align}
Overall normalization factors that do not affect the bridge conditions
will not be important below.

For two Ramond states, the leading ten-dimensional spin-field OPE is
part of the standard higher-point RNS spin-field technology
\cite{Haertl:2010qlb},
\begin{equation}
S_\alpha(z)S_\beta(0)
=
\frac{(C\gamma^\mu)_{\alpha\beta}}
{\sqrt{2}\,z^{3/4}}\,
\psi_\mu(0)
+\text{subleading terms},
\label{eq:app-spinfield-ope}
\end{equation}
while the superghost contribution is
\begin{equation}
e^{-\phi/2}(z)e^{-\phi/2}(0)
=
z^{-1/4}e^{-\phi}(0)+\cdots .
\label{eq:app-superghost-ope}
\end{equation}
Their product gives
\begin{equation}
V_{f,i}^{(-1/2)}(z)V_{f,j}^{(-1/2)}(0)
=
\frac{1}{2z}
\left(\chi_i\gamma^\mu\chi_j\right)
V_{\widehat{ij},g,\mu}^{(-1)}(0)
+O(1),
\label{eq:app-ff-local-fusion}
\end{equation}
where
\begin{equation}
k_{\widehat{ij}}=k_i+k_j .
\end{equation}
On the rectangular kinematic locus $s_{ij}=0$, the fused momentum is
null, and the singular term is therefore the massless fusion channel
$ff\to g$.

It is important that Eq.~\eqref{eq:app-ff-local-fusion} contains the
complete negative-power part of the fermion--fermion collision.  For
equal chirality, the next allowed spin-field channels carry higher-rank
fermionic operators and appear at higher powers of $z$.  Schematically,
\begin{equation}
S_\alpha(z)S_\beta(0)
\sim
z^{-3/4}(C\gamma^\mu)_{\alpha\beta}\psi_\mu
+
z^{1/4}(C\gamma^{\mu\nu\rho})_{\alpha\beta}
\psi_\mu\psi_\nu\psi_\rho
+\cdots .
\label{eq:app-spinfield-hierarchy}
\end{equation}
After multiplication by the universal superghost factor
$z^{-1/4}$, the vector channel is proportional to $z^{-1}$, whereas
the higher-rank channels begin at $O(1)$.  Hence
\begin{equation}
\operatorname{PP}_{z=0}
\left[
V_{f,i}^{(-1/2)}(z)V_{f,j}^{(-1/2)}(0)
\right]
=
\frac{1}{2z}
\left(\chi_i\gamma^\mu\chi_j\right)
V_{\widehat{ij},g,\mu}^{(-1)}(0),
\label{eq:app-ff-principal-part}
\end{equation}
where $\operatorname{PP}$ denotes the negative-power part of the local
expansion.  The bridge condition
\begin{equation}
\chi_i\gamma^\mu\chi_j=0
\end{equation}
therefore removes the complete fermion--fermion principal part.

The mixed bridge follows from the OPE of
$V_f^{(-1/2)}$ with $V_g^{(0)}$.  The contraction of
$\epsilon\cdot P$ with the fermion plane wave produces
$k_f\cdot\epsilon_g$, while the Lorentz-current term
$f_{\mu\nu}\psi^\mu\psi^\nu$ acts on the spin field and produces
$\chi_f\slashed f_g$.  Thus the potentially divergent mixed-channel
data are
\begin{equation}
k_f\cdot\epsilon_g,
\qquad
\chi_f\slashed f_g ,
\label{eq:app-fg-local-data}
\end{equation}
with the reversed collision giving the corresponding $gf$ bridge.

The NS--NS collision requires a slightly more careful regulated
statement.  Its potentially divergent terms have the schematic form
\begin{equation}
\begin{aligned}
V_{g,i}^{(0)}(z)V_{g,j}^{(0)}(0)
={}&
-\frac{s_{ij}\,
\epsilon_i\cdot\epsilon_j}{z^2}\,
\mathcal O_{ij}^{(0)}
+
\frac{1}{z}
\left[
(\epsilon_i\cdot k_j)\mathcal O_{ij}^{(i)}
+
(k_i\cdot\epsilon_j)\mathcal O_{ij}^{(j)}
+
s_{ij}\mathcal O_{ij}^{(s)}
\right]
+O(1),
\label{eq:app-gg-regulated-ope}
\end{aligned}
\end{equation}
where the $\mathcal O_{ij}^{(a)}$ are local operators regular in the
collision coordinate.  Terms containing products of the displayed
bridge contractions have been absorbed into the corresponding
operators and vanish under the same conditions.  The gluonic bridge
data are
\begin{equation}
\epsilon_i\cdot\epsilon_j,
\qquad
\epsilon_i\cdot k_j,
\qquad
k_i\cdot\epsilon_j .
\label{eq:app-gg-local-data}
\end{equation}
They remove the double pole and all simple-pole terms that would have
negative total scaling.  A term proportional to $s_{ij}/z$ may remain,
but on the regulated degeneration
\begin{equation}
s_{ij}=O(\tau),
\qquad
z=\sigma_{ij}=O(\tau),
\label{eq:app-gg-regulated-scaling}
\end{equation}
it is only $O(1)$.

The local fusion data therefore close on
\begin{equation}
ff\longrightarrow g,
\qquad
fg/gf\longrightarrow f,
\qquad
gg\longrightarrow g ,
\label{eq:app-local-fusion}
\end{equation}
and are controlled by the species-dependent bridges of
Eq.~\eqref{eq:species-bridge}.  For any cross pair, the uniform
regulated conclusion is
\begin{equation}
\left.
I_n
\right|_{
s_{ij}=O(\tau),\,
x_{ij}=0,\,
\sigma_{ij}=O(\tau)
}
=
O(1).
\label{eq:app-local-regulated-regularity}
\end{equation}
Equivalently, the bridge conditions remove all terms with negative total
$\tau$-scaling in the local collision expansion.  Equation
\eqref{eq:app-local-regulated-regularity}, rather than the stronger
requirement that every negative power of $\sigma_{ij}$ vanish
separately, is the local input needed in the general multi-pinch proof.

\subsection{Picture changing and arbitrary Ramond multiplicity}
\label{app:picture-changing}

For amplitudes with more than four Ramond insertions, additional
picture-changing operators (PCOs) are required to balance the total
picture number; see Ref.~\cite{Belopolsky:1997jz} for the general
picture-changing formalism.  The relevant point for the hidden-zero
analysis is that picture changing does not modify the fermionic
principal part associated with a pinched pair containing Ramond states.

We keep the PCOs as separate insertions and write the half-integrand
schematically as
\begin{equation}
I_n
=
\left\langle
V_i(z)V_j(0)\,
\prod_{A=1}^{N_{\rm PCO}} X(y_A)\,
\mathcal X
\right\rangle ,
\label{eq:app-pco-correlator}
\end{equation}
where $\mathcal X$ collects the remaining external vertices and
$z=\sigma_{ij}\to0$ is the active collision coordinate. For the local collision analysis, the PCO positions may be chosen at
finite separation from the collision, 
\begin{equation}
y_A\neq0,
\qquad
y_A-z=O(1),
\qquad
z\to0 .
\label{eq:app-pco-separation}
\end{equation}
Contractions involving $X(y_A)$ and either active vertex are therefore
analytic in $z$ and cannot generate additional negative powers of the
collision parameter.  The PCOs remain part of the spectator correlator
under the local fermionic fusion,
\begin{equation}
V_i(z)V_j(0)
\longrightarrow
\frac{1}{z}\,
(\text{fermionic bridge})\,
V_{\widehat{ij}}(0)
+O(1),
\label{eq:app-pco-local-fusion}
\end{equation}
without changing its singular coefficient.

It follows that the fermionic bridge analysis is independent of the
total number of Ramond insertions.  In particular, for an active
fermion--fermion pair,
\begin{equation}
\chi_i\gamma^\mu\chi_j=0
\qquad\Longrightarrow\qquad
\operatorname{PP}_{\sigma_{ij}=0}I_n=0
\label{eq:app-pco-principal-part}
\end{equation}
at arbitrary fermion multiplicity.  The analogous mixed fermionic
bridges remove the corresponding potentially divergent local
coefficients.  Increasing the number of external fermions changes the
spectator correlator and the number of required PCOs, but introduces no
new primitive fermionic data with negative total $\tau$-scaling at the
active collision.

As an independent check, one may picture-change one of the two active
Ramond vertices.  In the first-order matter representation we use
\begin{equation}
V_f^{(+1/2)}(\chi_j,k_j)
=
\mathcal N_+\,
e^{\phi/2}
\left[
P^\mu
+\frac14(k_j\cdot\psi)\psi^\mu
\right]
(\gamma_\mu\chi_j)^{\dot\beta}
S_{\dot\beta}\,
e^{ik_j\cdot X}.
\label{eq:app-plus-half-vertex}
\end{equation}
The external spinors obey
\begin{equation}
\slashed k_i\chi_i=0,
\qquad
\slashed k_j\chi_j=0,
\end{equation}
and we denote the fused momentum by
\begin{equation}
k_{\widehat{ij}}:=k_i+k_j .
\label{eq:app-mixed-fused-momentum}
\end{equation}

For the opposite-chirality spin fields appearing in the mixed-picture
collision, the terms relevant to the principal part are
\begin{equation}
S_\alpha(z)S_{\dot\beta}(0)
=
z^{-5/4}
\left[
\delta_{\alpha\dot\beta}
+\frac{z}{4}
(\gamma^{\rho\sigma})_{\alpha\dot\beta}
J_{\rho\sigma}(0)
+O(z^2)
\right],
\label{eq:app-opposite-spin-ope}
\end{equation}
where
\begin{equation}
J^{\mu\nu}=:\!\psi^\mu\psi^\nu\!: ,
\end{equation}
while the superghost contribution is
\begin{equation}
e^{-\phi/2}(z)e^{+\phi/2}(0)
=
z^{1/4}\bigl[1+O(z)\bigr].
\label{eq:app-mixed-ghost-ope}
\end{equation}
These terms exhaust the possible double and simple poles.

The $P^\mu$ part of
Eq.~\eqref{eq:app-plus-half-vertex} gives
\begin{equation}
\begin{aligned}
\operatorname{PP}_{P}
=
\mathcal N e^{ik_{\widehat{ij}}\cdot X}
\Bigg[
&
-\frac{k_{i,\mu}
\left(\chi_i\gamma^\mu\chi_j\right)}{z^2}
+\frac{
\left(\chi_i\gamma^\mu\chi_j\right)P_\mu}{z}
\nonumber\\
&-
\frac{1}{4z}
k_{i,\mu}
\chi_i\gamma^{\rho\sigma}\gamma^\mu\chi_j\,
J_{\rho\sigma}
\Bigg],
\end{aligned}
\label{eq:app-mixed-P-part}
\end{equation}
where $\mathcal N=\mathcal N_-\mathcal N_+$.  The apparent double pole
vanishes by the massless Dirac equation,
\begin{equation}
k_{i,\mu}
\left(\chi_i\gamma^\mu\chi_j\right)
=0 .
\end{equation}
Using the Clifford algebra,
\begin{equation}
k_{i,\mu}
\chi_i\gamma^{\rho\sigma}\gamma^\mu\chi_j
=
-2\left[
k_i^\rho
\left(\chi_i\gamma^\sigma\chi_j\right)
-
k_i^\sigma
\left(\chi_i\gamma^\rho\chi_j\right)
\right],
\label{eq:app-mixed-P-gamma}
\end{equation}
the remaining principal part becomes
\begin{equation}
\operatorname{PP}_{P}
=
\frac{\mathcal N}{z}\,
\left(\chi_i\gamma^\mu\chi_j\right)
\left[
P_\mu+(k_i\cdot\psi)\psi_\mu
\right]
e^{ik_{\widehat{ij}}\cdot X}.
\label{eq:app-mixed-P-result}
\end{equation}

The $(k_j\cdot\psi)\psi^\mu$ term gives the complementary contribution
\begin{equation}
\operatorname{PP}_{J}
=
\frac{\mathcal N}{z}\,
\left(\chi_i\gamma^\mu\chi_j\right)
(k_j\cdot\psi)\psi_\mu\,
e^{ik_{\widehat{ij}}\cdot X}.
\label{eq:app-mixed-J-result}
\end{equation}
Adding the two pieces gives
\begin{equation}
\operatorname{PP}_{z=0}
\left[
V_{f,i}^{(-1/2)}(z)
V_{f,j}^{(+1/2)}(0)
\right]
=
\frac{\mathcal N}{z}\,
\left(\chi_i\gamma^\mu\chi_j\right)
V_{\widehat{ij},g,\mu}^{(0)}(0),
\label{eq:app-mixed-picture-result}
\end{equation}
where
\begin{equation}
V_{\widehat{ij},g,\mu}^{(0)}
=
\left[
P_\mu+
(k_{\widehat{ij}}\cdot\psi)\psi_\mu
\right]
e^{ik_{\widehat{ij}}\cdot X}.
\label{eq:app-fused-zero-picture-gluon}
\end{equation}
No independent higher-rank spinor bilinear and no double pole survive.
The condition
\begin{equation}
s_{ij}=k_i\cdot k_j=0
\qquad\Longleftrightarrow\qquad
k_{\widehat{ij}}^2=0
\end{equation}
ensures that the fused vector state is massless.  Therefore
\begin{equation}
\chi_i\gamma^\mu\chi_j=0
\qquad\Longrightarrow\qquad
\operatorname{PP}_{z=0}
\left[
V_{f,i}^{(-1/2)}(z)
V_{f,j}^{(+1/2)}(0)
\right]
=0 ,
\end{equation}
confirming explicitly that picture changing introduces no new
fermionic bridge.

\subsection{Multi-pinch regularity}
\label{app:multipinch-regularity}

We now justify the simultaneous regulated $O(1)$ behavior of the RNS
half-integrand used in the main-text general multi-pinch proof.  Consider
a $p$-pinch sector \cite{Zhang:2024efe}
\begin{equation}
\mathfrak M_p
=
\left\{
(i_\alpha,j_\alpha)
\right\}_{\alpha=1}^{p},
\qquad
1\leq p\leq
\min\!\left(|L_m|,|R_m|\right),
\label{eq:app-multipinch-matching}
\end{equation}
with mutually row-- and column--disjoint pairs.  Introduce the local
collision coordinates
\begin{equation}
z_\alpha
:=
\sigma_{i_\alpha j_\alpha},
\qquad
z_\alpha=O(\tau),
\qquad
\alpha=1,\ldots,p ,
\label{eq:app-multipinch-coordinates}
\end{equation}
while the corresponding cross-pair invariants obey
\begin{equation}
s_{i_\alpha j_\alpha}=O(\tau).
\label{eq:app-multipinch-invariants}
\end{equation}
Because the matching pairs are disjoint, the corresponding collision
regions are mutually separated at leading order.

Consider first one pair $(i_\alpha,j_\alpha)$.  All remaining
insertions, including the other $p-1$ colliding pairs, may be collected
into a spectator operator $\mathcal X_\alpha$,
\begin{equation}
I_n
=
\left\langle
V_{i_\alpha}(z_\alpha)
V_{j_\alpha}(0)\,
\mathcal X_\alpha
\right\rangle .
\label{eq:app-multipinch-spectator}
\end{equation}
The local analysis of
Secs.~\ref{app:rns-local-opes} and
\ref{app:picture-changing} then gives
\begin{equation}
\left.
I_n
\right|_{
s_{i_\alpha j_\alpha}=O(\tau),\,
x_{i_\alpha j_\alpha}=0,\,
z_\alpha=O(\tau)
}
=
O(1)
\label{eq:app-one-pinch-regularity}
\end{equation}
with respect to the active collision, independently of the species and
number of the spectator insertions.  In the fermion--fermion channel
this follows from the disappearance of the complete principal part.  In
the mixed channels the bridge removes the potentially divergent local
coefficients.  In the gluon--gluon channel terms such as
$s_{i_\alpha j_\alpha}/z_\alpha$ may remain, but have zero total
$\tau$-weight and are therefore $O(1)$.

Since the collision regions are mutually disjoint, the corresponding
local OPEs may be applied independently.  A general term in the joint
collision expansion may contain negative powers of individual
$z_\alpha$, but each such power is accompanied by sufficient powers of
the regulated cross-pair invariants to give nonnegative total
$\tau$-weight after the full bridge conditions are imposed.  Thus no
term with negative total $\tau$-scaling survives on the multi-pinch
sector, and consequently
\begin{equation}
\left.
I_n
\right|_{\mathfrak M_p,H_{L_m|R_m}}
=
O(1)
\label{eq:app-multipinch-regularity}
\end{equation}
on every $p$-pinch branch.

This establishes the only additional worldsheet ingredient required
for the general multi-pinch extension.  Multiple pinches introduce no
new fermionic fusion data: they simply place several mutually disjoint
local collisions on the same scattering-equation solution.  Together
with the scattering-equation scaling and Parke--Taylor regularity used
in the main-text worldsheet proof, this reproduces the general
multi-pinch suppression and hence Eq.~\eqref{eq:main-hidden-zero}.

\section{Graviton, gravitino, and EYM components}
\label{app:gravity-components}

We collect here a unified description of the gravitational and
gauge--gravity components discussed in the main text.  The essential
point is that the two factorized half-integrands carry independent
external-state data, while the color sector enters through the
Parke--Taylor factor.  This allows gluons, gluinos, gravitons, and
gravitinos to be treated within a common single-trace construction, with
the examples of the main text arising as simple restrictions of the
external-state content.

We choose the factorized state assignment
\begin{equation}
g:\ \epsilon,
\qquad
f:\ \chi,
\qquad
h:\ \epsilon_\mu\widetilde\epsilon_\nu,
\qquad
\psi_{3/2}:\ \epsilon_\mu\widetilde\chi_\alpha,
\qquad
\slashed{\epsilon}\,\widetilde\chi=0 .
\label{eq:app-unified-states}
\end{equation}
For the last line the gamma-trace condition projects the factorized
NS--R vector-spinor onto the spin-$3/2$ gravitino sector.  The corresponding
state content seen by the two half-integrands is
\begin{equation}
\begin{array}{c|cc}
\text{state}
&
I_n
&
\widetilde I_{\rm grav}
\\ \hline
g & g & \varnothing \\
f & f & \varnothing \\
h & g & g \\
\psi_{3/2} & g & f
\end{array}
\label{eq:app-unified-state-table}
\end{equation}
Thus $I_n$ carries the untilded data $(\epsilon,\chi)$ of all external
states, whereas $\widetilde I_{\rm grav}$ carries the tilded data
$(\widetilde\epsilon,\widetilde\chi)$ only for gravitationally dressed
states.

A general single-trace component therefore has the schematic CHY
representation \cite{Cachazo:2014xea,Stieberger:2016lng}
\begin{equation}
\mathcal A_n^{\rm st}
=
\int d\mu_n\,
\operatorname{PT}(\mathbb I_{\rm c})\,
I_n(\epsilon,\chi)\,
\widetilde I_{\rm grav}
(\widetilde\epsilon,\widetilde\chi),
\label{eq:app-unified-single-trace-chy}
\end{equation}
where $\mathbb I_{\rm c}$ denotes the single color trace.

The bridge assignment can be stated without enumerating all physical
pairs separately.  In either copy, a given external leg appears as a
gluon, a fermion, or an empty state according to
Eq.~\eqref{eq:app-unified-state-table}.  For two nonempty endpoints the
bridge is precisely the corresponding gauge-theory bridge of
Eq.~\eqref{eq:species-bridge}.  When one endpoint is empty, only the
momentum--polarization contraction associated with a vector endpoint can
remain.  Explicitly,
\begin{equation}
\begin{array}{c|ccc}
x(X,Y)
&
Y=\varnothing
&
Y=g
&
Y=f
\\ \hline
X=\varnothing
&
\varnothing
&
\{k_i\!\cdot\!\epsilon_j\}
&
\varnothing
\\[1mm]
X=g
&
\{\epsilon_i\!\cdot k_j\}
&
\{\epsilon_i\!\cdot\!\epsilon_j,\,
  \epsilon_i\!\cdot k_j,\,
  k_i\!\cdot\epsilon_j\}
&
\{\epsilon_i\!\cdot k_j,\,
  \slashed f_i\chi_j\}
\\[1mm]
X=f
&
\varnothing
&
\{k_i\!\cdot\epsilon_j,\,
  \chi_i\slashed f_j\}
&
\{\chi_i\gamma^\mu\chi_j\}
\end{array}
\label{eq:app-universal-bridge-table}
\end{equation}
where $X$ and $Y$ denote the effective species carried by legs $i$ and
$j$, respectively, in the relevant half-integrand.  The symbol
$\varnothing$ denotes the absence of polarization or spinor data in that
copy; the external momentum is still present.  An empty bridge set
therefore imposes no additional condition.

For the tilded copy, exactly the same table applies with
\begin{equation}
\epsilon\rightarrow\widetilde\epsilon,
\qquad
\chi\rightarrow\widetilde\chi,
\qquad
f^{\mu\nu}\rightarrow\widetilde f^{\mu\nu}.
\label{eq:app-tilded-replacement}
\end{equation}
Equations~\eqref{eq:app-unified-state-table} and
\eqref{eq:app-universal-bridge-table} therefore determine the complete
bridge data for any pair of physical external states.

For example,
\begin{equation}
(\psi_{3/2},h):
\qquad
I_n:\ (g,g),
\qquad
\widetilde I_{\rm grav}:\ (f,g),
\label{eq:app-psi-h-example}
\end{equation}
so the untilded copy carries the bosonic bridge while the tilded copy
carries the fermion--gluon bridge.  Similarly,
\begin{equation}
(f,h):
\qquad
I_n:\ (f,g),
\qquad
\widetilde I_{\rm grav}:\ (\varnothing,g),
\label{eq:app-f-h-example}
\end{equation}
so the tilded copy contributes only the restricted bridge
$k_f\cdot\widetilde\epsilon_h$.

We now specify the rectangular partition.  Choose two distinguished
external legs $a,b$, and divide all remaining legs into two nonempty
sets
\begin{equation}
A\sqcup B
=
\{1,\ldots,n\}\setminus\{a,b\},
\qquad
A,B\neq\varnothing .
\label{eq:app-single-trace-partition}
\end{equation}
For every $(i,j)\in A\times B$, impose
\begin{equation}
s_{ij}=0,
\qquad
x_{ij}=0,
\qquad
\widetilde x_{ij}=0,
\label{eq:app-single-trace-bridge-conditions}
\end{equation}
where $x_{ij}$ and $\widetilde x_{ij}$ are read from the two copies
above.  These bridge conditions remove all contributions with negative total
$\tau$-scaling from both $I_n$ and $\widetilde I_{\rm grav}$ on every
pinched cross pair.

For a single-trace component, the remaining requirement is the
regularity of the Parke--Taylor factor.  The condition can be stated
uniformly: the rectangular partition must be chosen so that no
Parke--Taylor edge directly connects a leg in $A$ to a leg in $B$.
Equivalently, every $A$--$B$ transition along the color trace
$\mathbb I_{\rm c}$ must pass through one of the distinguished legs
$a$ or $b$.

When both distinguished legs belong to the color trace, this is realized
by a decomposition
\begin{equation}
\mathbb I_{\rm c}
=
(a,A',b,B'),
\label{eq:app-pt-compatible-ordering}
\end{equation}
with the remaining gravitational states distributed arbitrarily between
the two sides.  If one or both distinguished legs lie outside the color
trace, the same criterion simply requires all traced states not separated
by $a$ and $b$ to remain on the same side of the rectangular partition.

Under this condition,
\begin{equation}
\operatorname{PT}(\mathbb I_{\rm c})=O(1)
\label{eq:app-single-trace-regularity}
\end{equation}
on every multi-pinch degeneration associated with $A\times B$.

Together with the bridge conditions,
\begin{equation}
I_n=O(1),
\qquad
\widetilde I_{\rm grav}=O(1),
\qquad
\operatorname{PT}(\mathbb I_{\rm c})=O(1).
\end{equation}
For a $p$-pinch sector the universal scattering-equation Jacobian gives
\begin{equation}
(\det{}'\Phi)^{-1}=O(\tau^p),
\end{equation}
so every such sector is suppressed as $O(\tau^p)$.  This yields the
corresponding single-trace hidden zero.

At the level of the factorized CHY integrand, the YM,
single-trace EYM, and gravity structures are therefore
related schematically by
\begin{equation}
\operatorname{PT}\,I_n
\quad\longrightarrow\quad
\operatorname{PT}\,I_n\,\widetilde I_{\rm grav}
\quad\longrightarrow\quad
I_n\,\widetilde I_n .
\label{eq:app-ym-eym-gravity-interpolation}
\end{equation}
The arrows refer to the factorized CHY structure rather than to a
continuous deformation of a fixed external-state component.  Pure
gluon--gluino amplitudes retain the first structure; single-trace
gauge--gravity amplitudes realize the second; and fully gravitational
amplitudes contain the two complete half-integrands.

\end{document}